\documentclass[12pt]{article}
\usepackage{amssymb,amsmath,epsfig}
\usepackage{lscape}

\begin{document}
\title{\bf Cosmological reconstruction and energy bounds in $f(R,R_{\alpha
\beta}R^{\alpha\beta},\phi)$ gravity}
\author{M. Zubair\thanks{mzubairkk@gmail.com; drmzubair@ciitlahore.edu.pk}
and Farzana Kousar\thanks{farzana.kouser83@gmail.com}\\
Department of Mathematics, COMSATS\\
Institute of Information Technology Lahore, Pakistan.}

\date{}

\maketitle

\begin{abstract}
We discuss the cosmological reconstruction of
$f(R,R_{\alpha\beta}R^{\alpha\beta},\phi)$ (where $R$,
$R_{\alpha\beta}R^{\alpha\beta}$ and $\phi$ represents the Ricci
scalar, Ricci invariant and scalar field) corresponding to power law
and de Sitter evolution in the framework of FRW universe model. We
derive the energy conditions for this modified theory which seem to
be more general and can be reduced to some known forms of these
conditions in general relativity, $f(R)$ and $f(R,\phi)$ theories.
We have presented the general constraints in terms of recent values
of snap, jerk, deceleration and Hubble parameters. The energy bounds
are analyzed for reconstructed as well as known models in this
theory. Finally, the free parameters are analyzed comprehensively.
\end{abstract}

\section{Introduction}

In current cosmic picture dark energy (DE) is introduced as an
effective characteristic which tends to accelerate the expansion in
universe. Modified theories have achieved significant attention to
explore the effect of cosmic acceleration \cite{1}. These models
have been developed to distinguish the source of DE as modification
to the Einstein Hilbert action. Some modified theories of gravity
are $f(R)$ gravity with Ricci scalar $R$ \cite{2}, $f(T)$ gravity
with torsion scalar $T$ \cite{3}, Gauss-Bonnet gravity with
$\mathcal{G}$ invariant \cite{4}, $f(R,\mathcal{T})$ gravity with
$\mathcal {T}$ as the trace of stress-energy tensor \cite{5},
$f(R,\mathcal{T},R_{\mu\nu})T^{\mu\nu}$ \cite{5a} and
$f(R,\mathcal{G})$ gravity that contains both $R$ and $\mathcal{G}$
\cite{6} etc. The acceleration of the expanding universe can be
explored by these theories through their comparable invariants.

To generalize Einstein's theory of general relativity (GR), there is
a vast literature on relativistic theories that reduce to GR in the
proper limitations. An especially attractive class of these
generalizations are the fourth-order theories. These theories were
initially considered by Eddington in early 1920's \cite{7}.
Whatever the inspiration to examine the generalized fourth-order
theories, it is necessary to understand their weak-field limit, and
these limits confirm the increasing behavior of these theories in
observational data.

Generally a fourth order theory of gravity is obtained by adding
$R_{ab} R^{ab}$ and $R_{abcd}R^{abcd}$ in the standard Einstein
Hilbert action \cite{8,9}. However, it is now established that we
can ignore the $R_{abcd} R^{abcd}$ term if we use the Gauss Bonnet
theorem \cite{10}. About a half century back, Brans and Dicke (BD)
\cite{11} presented the scalar-tensor theory of gravitation, which
is still popular and have received great interest in cosmological
dynamics as a replacement to dark matter and dark energy. The
motivation behind the BD theory was Mach idea \cite{12} to present a
varying gravitational constant in general relativity. Amongst the
alternative theories to Einstein's gravity, the simplest and well
known is Brans-Dicke theory. In this theory, the gravitational
constant has been taken to be inversely proportional to the scalar
field $\phi$. The BD theory may be presented as a generalization of
$f(R)$ theory with $f'(R)=F(R)=\phi R$ \cite{2}.

In modified theories, cosmological reconstruction is one of the
important prospects in cosmology. In $f(R)$ gravity, the
reconstruction scheme has been used in different contexts to explain
the conversion of matter dominated era to DE phase. This can ne
examined by considering the known cosmic evolution and the field
equations are used to calculate particular form of Lagrangian which
can reproduce the given evolution background. In these theories the
existence of exact power law solutions for FRW spacetime has been
examined. In \cite{13,14,15} people have reconstructed $f(R,T)$
gravity models by employing various cosmological scenarios. Nojiri
et al. developed $f(R)$ gravity models \cite{16}, which were further
applied to $f(R, G)$ and modified Gauss-Bonnet theories \cite{17}.
To reconstruct $f(R)$ gravity models, Carloni et al. \cite{18} has
established a new technique by using the cosmic parameters instead
of using scale factor.

Energy conditions are necessary to study the singularity theorems moreover
the theorems related to black hole thermodynamics. For example, the well
known Hawking-Penrose singularity theorems \cite{19} invoke the null energy
condition (NEC) as well as strong energy condition (SEC). The  violation  of
(SEC) allows to observe the accelerating expansion, and null energy
conditions (NEC) are involved in proof of second law of black hole
thermodynamics.

Energy conditions have been explored in different contexts like
$f(T)$ theory \cite{20,20a}, $f(R)$ gravity \cite{21} and $f(G)$
theory \cite{22}, Brans-Dicke theory \cite{23}. Further the energy
conditions of a very generalized second order scalar tensor gravity
have been discussed by Sharif and Saira \cite{24}. Sharif and Zubair
have examined these conditions for $f(R,T)$ gravity \cite{13} and
for $f(R,T,R_{\mu\nu}T^{\mu\nu})$ gravity \cite{26} which involves
the nonminimal coupling between the Ricci tensor and energy-momentum
tensor. Saira and Zubair \cite{27} have discussed these conditions
for $F(T,T_{G})$ having term $T$ torsion invariant along with
$T_{G}$, equivalence of Gauss-Bonnet term and teleparallel.

In this paper we are interested to develop some cosmic models
coherent with the recent observational data in the vicinity of
generalized scalar tensor theories. We present the energy conditions
in $f(R,R_{\alpha\beta}R^{\alpha\beta},\phi)$ gravity utilizing FRW
universe model with perfect fluid matter and developed some constraints
on free parameters on reconstructed as well as well known models. The
paper is arranged in the following pattern: In next section, we are
providing a general introduction of $f(R,R_{\alpha\beta}R^{\alpha
\beta},\phi)$ gravity. In section 3 we have defined the basic
expressions of energy conditions and then derive the energy
conditions of $f(R,R_{\alpha \beta}R^{\alpha\beta},\phi)$ gravity
using deceleration, jerk and snap parameters. Section 4 is devoted
to the reconstruction of models in $f(R,R_{\alpha
\beta}R^{\alpha\beta},\phi)$ gravity and energy bounds of these
models and in section 5 we have derived the energy conditions of
some known $f(R,\phi)$ models. In section 6, we sum up our
conclusion.

\section{Scalar Tensor fourth Order Gravity}

The $f(R, R_{\alpha\beta}R^{\alpha\beta},\phi)$ gravity has an
interesting prospect among the more general scalar tensor theories
and its action is of the form \cite{28},
\begin{equation}\label{18}
S_{m}=\int d^{4}x \sqrt{-g} \left[\frac{1}{\kappa^2}\left(f\left(R,R_{\alpha
\beta}R^{\alpha\beta},\phi\right)+\omega(\phi)\phi_{;\alpha}\phi^{;\alpha}
\right)+\mathcal{L}_{m}\right],
\end{equation}
where $f$ is an unspecified function of the Ricci scalar, the curvature
invariant and the scalar field denoted by $R$, $R_{\alpha\beta}R^{\alpha
\beta}\equiv Y$ and $\phi$  (where $R_{\alpha\beta}$ is the Ricci tensor).
The $\mathcal{L}_{m}$ is the matter Lagrangian density, $\omega$ is a generic
function of the scalar field $\phi$, $g$ is the determinant of the metric
tensor $g_{\mu\nu}$.

In the metric approach, by varying the action (\ref{18}) with
respect to $g_{\mu\nu}$ the field equations are obtained as
\begin{eqnarray}\label{1}
&&\nonumber f_{R}R_{\mu\nu}-\frac{1}{2}\left(f+\omega(\phi) \phi_{;\alpha}
\phi^{;\alpha}\right)g_{\mu\nu}-f_{R;\mu\nu}+g_{\mu\nu}\Box f_{R}+2f_{Y}
R_{\mu}^{\alpha}R_{\alpha\nu}\\
&&-2[f_{Y}R^{\alpha}_{(\mu}]_{;\nu)\alpha}+\Box [f_{Y}R_{\mu\nu}]+[f_{Y}
R_{\alpha\beta}]^{;\alpha\beta}g_{\mu\nu}+\omega(\phi)\phi_{;\mu}\phi_{;\nu}
=\kappa^2 T_{\mu\nu},
\end{eqnarray}
where $\Box=g^{\mu\nu}\nabla_{\mu}\nabla_{\nu}$ and $\kappa^2\equiv 8 \pi G$.
We consider the flat FRW universe model with $a(t)$ as scale factor given by
\begin{equation}\label{8}
ds^{2}=dt^{2}-a^{2}(t)\left(dx^{2}+dy^{2}+dz^{2}\right),
\end{equation}
The gravitational field equations corresponding to perfect fluid as
matter content, are given by
\begin{eqnarray}
\nonumber\noindent\kappa^2 \rho&=&-3\left(\dot{H}+H^{2}\right)f_{R}+3H
\partial_{t}f_{R}-\frac{1}{2}\left(f-\omega(\phi)\dot{\phi}^{2}\right)-6H\left
(2\dot{H}+3H^{2}\right)\\
&&\times\partial_{t}f_{Y}+\left(114\dot{H}H^{2}+24\dot{H}^{2}+42H^{4}\right)
f_{Y}, \label{2}\\
\nonumber\kappa^2 p&=&\frac{1}{2}\left(f+\omega(\phi)\dot{\phi}^{2}\right)
+\left(\dot{H}+3H^{2}\right)f_{R}-2H\partial_{t}f_{R}-\partial_{tt}f_{R}+4H
\left(\dot{H}\right.\\
\nonumber&&\left.+3H^{2}\right)\partial_{t}f_{Y}+\left(4\dot{H}+6H^{2}\right)
\partial_{tt}f_{Y}+\left(4\dddot{H}+20\ddot{H}H+10\dot{H}H^{2}\right.\\
&&\left.+16\dot{H}^{2}-18H^{4}\right)f_{Y}.\label{3}
\end{eqnarray}
The field equation (\ref{1}) can be rearranged in the following form
\begin{equation}\label{4}
G_{\mu\nu}=R_{\mu\nu}-\frac{1}{2}Rg_{\mu\nu}=T_{\mu\nu}^{eff},
\end{equation}
which is similar to the standard field equations in GR. Here
$T_{\mu\nu}^{eff}$, the effective energy-momentum tensor in $f( R, Y, \phi)$
gravity is defined as
\begin{eqnarray}\label{5}
\nonumber&&T_{\mu\nu}^{eff}=\frac{1}{f_{R}}\bigg[\kappa^2 T_{\mu\nu}+\frac{1}{2}
\left(f+\omega(\phi)\phi_{;\alpha}\phi^{;\alpha}-R f_{R}\right)g_{\mu \nu}
+f_{R;\mu\nu}-g_{\mu\nu}\Box f_{R}\\
&&\nonumber-2f_{Y}R_{\mu}^{\alpha}R_{\alpha\mu}+2[f_{Y}
R^{\alpha}_{(\mu}]_{;\nu)\alpha}-\Box [f_{Y}R_{\mu\nu}]-[f_{Y}
R_{\alpha\beta}]^{;\alpha\beta}g_{\mu\nu}-\omega(\phi) \phi_{;\mu}
\phi_{;\nu}~\bigg].
\end{eqnarray}
One can define the effective energy density and pressure of the form
\begin{eqnarray}\label{6}
\nonumber&&\rho_{eff}=\frac{1}{f_{R}}\left[\kappa^2\rho+\frac{1}{2}\left(f
-\omega(\phi)\dot{\phi}^{2}\right)+3\left(\frac{\dot{a}^{2}}{a^{2}}
+\frac{\ddot{a}}{a}\right)f_{R}-3\frac{\dot{a}}{a}\partial_{t}f_{R}+6
\left(\frac{\dot{a}^{3}}{a^{3}}\right.\right.\\
&&\left.\left.+2\frac{\dot{a}\ddot{a}}{a^{2}}\right)\partial_{t}f_{Y}
+\left(24 \frac{\ddot{a}^{2}}{a^{2}}-66\frac{\dot{a}^{2}\ddot{a}} {a^{3}}
+48\frac{\dot{a}^{4}}{a^{4}}\right)f_{Y}\right],
\end{eqnarray}
and
\begin{eqnarray}\label{7}
\nonumber&&p_{eff}=\frac{1}{f_{R}}\left[\kappa^2 p-\frac{1}{2}\left(f+\omega(\phi)
\dot{\phi}^{2}\right)-3\left(\frac{\ddot{a}}{a}+\frac{\dot{a}^{2}}{a^{2}}\right)
f_{R}+2\frac{\dot{a}}{a}\partial_{t}f_{R}+\partial_{tt}f_{R}\right.\\
\nonumber&&\left.-2\left(\frac{\ddot{a}}{a}+2\frac{\dot{a}^{2}}{a^{2}}\right)
\partial_{tt}f_{Y}-4\left(\frac{\dot{a}\ddot{a}}{a^{2}} +2\frac{\dot{a}^{3}}
{a^{3}}\right)\partial_{t}f_{Y}-\left(4\frac{\ddddot{a}}{a}+4\frac{\dot{a}
\dddot{a}}{a^{2}}-34\frac{\dot{a}^{2}\ddot{a}}{a^{3}}\right.\right.\\
&&\left.\left.-4\frac{\ddot{a}^{2}}{a^{2}}-4\frac{\dot{a}^{4}}{a^{4}}\right)
f_{Y}\right].
\end{eqnarray}

\section{Energy Conditions}

Energy conditions have an important role in GR, and also have useful
applications in modified theories of gravity. In the context of GR,
these constraints help to constrain the possible choices of matter
contents. Four types of energy conditions are developed in GR by
applying a geometrical result known as Raychaudhuri equation
\cite{19}. These conditions are known as null energy condition
(NEC), weak energy condition (WEC), strong energy condition (SEC)
and dominant energy condition (DEC).

In a spacetime manifold, the temporal evolution of expansion scalar is
described as Raychaudhuri equation given by,
\begin{equation}\label{20}
\frac{d\theta}{d\tau}=-\frac{1}{3}\theta^{2}-\sigma_{\mu\nu}\sigma^{\mu\nu}
+\omega_{\mu\nu}\omega^{\mu\nu}-R_{\mu\nu}u^{\mu}u^{\nu},
\end{equation}
\begin{equation}\label{21}
\frac{d\theta}{d\tau}=-\frac{1}{3}\theta^{2}-\sigma_{\mu\nu}\sigma^{\mu\nu}
+\omega_{\mu\nu}\omega^{\mu\nu}-R_{\mu\nu}k^{\mu}k^{\nu},
\end{equation}
where $R^{\mu\nu}$, $\sigma_{\mu\nu}$, $\omega^{\mu\nu}$ are Ricci tensor,
shear tensor and rotation, and the tangent vectors to timelike and null-like
curves in the congruence are represented by $u^{\mu}$ and $k^{\mu}$. The
interesting aspect of gravity makes the congruence geodesic convergent and
leads to the condition $\frac{d\theta}{d\tau}<0$. By ignoring the second
-order terms and integrating, the Raychaudhuri equation implies that $\theta=
-\tau R_{\mu\nu}u^{\mu}u^{\nu}$ and $\theta=-\tau R_{\mu\nu} k^{\mu}k^{\nu}$.
It further leads to the inequalities
\begin{equation}\label{22}
R_{\mu\nu}u^{\mu}u^{\nu}\geq0,~~~R_{\mu\nu}k^{\mu}k^{\nu}\geq0,
\end{equation}
These inequalities can be written as a linear combination of energy-momentum
tensor and its trace by the inversion of the gravitational field equations as
follows:
\begin{equation}\label{23}
\left(T_{\mu\nu}-\frac{T}{2}g_{\mu\nu}\right)u^{\mu}u^{\nu}\geq0,~~~
\left(T_{\mu\nu}-\frac{T}{2}g_{\mu\nu}\right)k^{\mu}k^{\nu}\geq0.
\end{equation}
In case of perfect fluid with density $\rho$ and pressure $p$, these
inequalities gives NEC, WEC, SEC, and DEC defined by:
\begin{eqnarray}\label{24}
\nonumber NEC&:&\rho+p\geq0,\\
\nonumber WEC&:&\rho\geq0,~~\rho+p\geq0,\\
\nonumber SEC&:&\rho+p\geq0,~~\rho+3p\geq0,\\
DEC&:&\rho\geq0,~~\rho\pm p\geq0.
\end{eqnarray}
In modified theories of gravity, assuming that the total matter contents act
like perfect fluid, these conditions can be determined by interchanging $\rho$
with $\rho_{eff}$ and $p$ with $p_{eff}$.\\
Energy conditions for scalar tensor fourth order gravity are:
\begin{eqnarray}
\nonumber&&\textbf{NEC:}~~\rho_{eff}+p_{eff}=\frac{1}{f_{R}}\left[\kappa^2\left
(\rho+p\right)-\omega(\phi)\dot{\phi}^{2}+\partial_{tt}f_{R}-H\partial_{t}
f_{R}-2\big(2\dot{H}\right.\\
\nonumber&&\left.+3H^{2}\big)\partial_{tt}f_{Y}+\left(8\dot{H}H+6H^{3}\right)
\partial_{t}f_{Y}-\left(4\dddot{H}+20H\ddot{H}+28\dot{H}H^{2}\right.\right.\\
&&\left.\left.+40\dot{H}^{2}\right)f_{Y}\right],\label{9}\\
\nonumber&&\textbf{WEC:}~~\rho_{eff}=\frac{1}{f_{R}}\left[\kappa^2\rho+\frac{1}{2}
\left(f-\omega(\phi)\dot{\phi}^{2}-R f_{R}\right)-3H\partial_{t}f_{R}+6H
\left(2\dot{H}\right.\right.\\
&&\left.\left.+3H^{2}\right)\partial_{t}f_{Y}-\left(18\dot{H}H^{2}
+24\dot{H}^{2}+18H^{4}\right)f_{Y}\right],\label{10}
\end{eqnarray}
\begin{eqnarray}
\nonumber&&\textbf{SEC:}~~\rho_{eff}+3 p_{eff}=\frac{1}{f_{R}}\left[\kappa^2\left
(\rho+3p\right)-f-2\omega(\phi)\dot{\phi}^{2}+R f_{R}+3H\partial_{t}f_R\right.\\
\nonumber&&\left.+3\partial_{tt}f_{R}-6\left(2\dot{H}+3H^{2}\right)
\partial_{tt} f_{Y}-18H^{3}\partial_{t}f_{Y}-\left(12\dddot{H}+60\ddot{H}H
\right.\right.\\
&&\left.\left.+48\dot{H}H^{2}+72\dot{H}^{2}-36H^{4}\right)f_{Y}\right],
\label{11}\\
\nonumber&&\textbf{DEC:}~~\rho_{eff}-p_{eff}=\frac{1}{f_{R}}\left[\kappa^2\left
(\rho-p\right)+f-R f_{R}-\partial_{tt}f_{R}-5H\partial_{t}f_{R}\right.\\
\nonumber&&\left.+2\left(2\dot{H}+3H^{2}\right)\partial_{tt}f_{Y}
+\left(16\dot{H}H+30H^{3}\right)\partial_{t}f_{Y}+\left(4\dddot{H}+20H\ddot{H}
\right.\right.\\
&&\left.\left.-8\dot{H}H^{2}-8\dot{H}^{2}-36H^{4}\right)f_{Y}\right].\label{12}
\end{eqnarray}
Inequalities (\ref{9})-(\ref{12}) represent the null, weak, strong and
dominant energy conditions in the context of $f(R,Y,\phi)$ gravity for FRW
spacetime.

We define the Ricci scalar and its derivatives in terms of deceleration, jerk
and snap parameters as \cite{29,30}
\begin{equation}\label{12a}
R=-6H^{2}(1-q),~~\dot{R}=-6H^{3}(j-q-2),~~\ddot{R}=6H^{4}(s+q^{2}+8q+6),
\end{equation}
where
\begin{equation}\label{13}
q=-\frac{1}{H^2}\frac{\ddot{a}}{a},~~j=\frac{1}{H^3}\frac{\dddot{a}}{a},
~~s=\frac{1}{H^4}\frac{\ddddot{a}}{a},
\end{equation}
and express Hubble parameter and its time derivatives in terms of these
parameters as \cite{26,27}
\begin{eqnarray}\label{14}
\nonumber&&H=\frac{\dot{a}}{a},~~\dot{H}=-H^2\left(1+q\right),~~\ddot{H}=
\left(j+3q+2\right)H^3,\\
&&\dddot{H}=H^4\left(s-4j-12q-3q^2-6\right).
\end{eqnarray}
Using the above definitions, the energy conditions (\ref{9})-(\ref{12}) can be
rewritten as
\begin{eqnarray}
\nonumber&&\textbf{NEC:}~~\kappa^2\left(\rho+p\right)-\omega(\phi)\dot{\phi}^2
-6H^4\left(s-j+(q+1)(q+8)\right)f_{RR}+\left\{\ddot{Y}\right.\\
\nonumber&&\left.-H\dot{Y}+12H^6\left(s(1-2q)+j\left(1+4q\right)+\left(q+1
\right)\left(-2q^2-17q+4\right)\right)\right\}\\
\nonumber&&\times f_{RY}+\left(\ddot{\phi}-H\dot{\phi}\right)f_{R\phi}-2H^{2}
\left(\ddot{Y}+H\dot{Y}\right)-2q\left(\ddot{Y}-2H\dot{Y}\right)f_{YY}-2H^2\\
\nonumber&&\times\left\{\left(\ddot{\phi}+H\dot{\phi}\right)
-2q\left(\ddot{\phi}-2H\dot{\phi}\right)\right\}f_{Y\phi}+36H^{6}\left(j-q
-2\right)^{2}f_{RRR}-12H^3\\
\nonumber&&\times\left(j-q-2\right)\left\{\dot{Y}+6H^5\left(1-2q\right)
\left(j-q-2\right)\right\}f_{RRY}-12\dot{\phi}H^3\left(j-q-2\right)
\end{eqnarray}
\begin{eqnarray}
\nonumber&&\times f_{RR\phi}+\dot{Y}\left\{\dot{Y}+24H^5(1-2q)(j-q-2)\right\}
f_{RYY}+\dot{\phi}^2f_{R\phi\phi}+2\dot{\phi}\left\{\dot{Y}\right.\\
\nonumber&&\left.+12 H^5(1-2q)(j-q-2)\right\}f_{RY\phi}+\dot{Y}\left(\dot{Y}
f_{YYY}+\dot{\phi}f_{YY\phi}\right)+\dot{\phi}\left(\dot{\phi}f_{Y\phi\phi}
\right.\\
&&\left.+\dot{Y} f_{YY\phi}\right)-4H^4\left(s+j+7q^2+16q+7\right)f_{Y}\geq0,
\label{17}\\
\nonumber&&\textbf{WEC:}~~\kappa^2\rho+\frac{1}{2}\left(f-\omega(\phi)\dot{\phi}^2
\right)-\frac{1}{2}R f_{R}+18H^4\left(j-q-2\right)\left\{f_{RR}-2H^2\right.\\
\nonumber&&\left.\times\left(1-2q\right)f_{RY}\right\}-3H\left(\dot{Y}f_{RY}
+\dot{\phi}f_{R\phi}\right)+6H^3\left(1-2q\right)\left(\dot{Y}f_{YY}+\dot{\phi}
f_{Y\phi}\right)\\
&&-6H^4\left(4q^2+5q+4\right)f_{Y}\geq0,\label{16}\\
\nonumber&&\textbf{SEC:}~~\kappa^2\left(\rho+3p\right)-f-2\omega(\phi)
\dot{\phi}^{2}+R f_{R}-6H^{4}\left(2s+2j-6q^{2}+14q\right)\\
\nonumber&&+17f_{Y}-18H^{4}\left(s+j+q^2+7q+4\right)f_{RR}+3\left\{H\dot{Y}
+\ddot{Y}+12H^6\left(1-2q\right)\right.\\
\nonumber&&\left.\left(s+8q+q^2+6\right)+36H^{6}\left(j-q-2\right)\right\}
f_{RY}+3\left(H\dot{\phi}+\ddot{\phi}\right)f_{R\phi}-6H^2\times\\
\nonumber&&\left(\left(1-2q\right)\ddot{Y}+3H\dot{Y}\right)f_{YY}-6H^2\left
(\left(1-2q\right)\ddot{\phi}+3H\dot{\phi}\right) f_{Y\phi}+108H^{6}\times\\
\nonumber&&\left(j-q-2\right)^2f_{RRR}-36H^3\left(j-q-2\right)\left(\dot{Y}
+6H^5\left(1-2q\right)\left(j-q-2\right)\right)\times\\
\nonumber&&f_{RRY}-36H^3\left(j-q-2\right)\left\{\dot{Y}+6H^5\left(1-2q\right)
\left(j-q-2\right)\right\}f_{RRY}-36H^3\times\\
\nonumber&&(j-q-2)\dot{\phi}f_{RR\phi}+3\left\{\dot{Y}^2+24H^5\dot{Y}(1-2q)
(j-q-2)\right\}f_{RYY}+3\dot{\phi}^2f_{R\phi\phi}\\
\nonumber&&+6\left\{\dot{\phi}\dot{Y}+12H^5\dot{\phi}(1-2q)(j-q-2)\right\}
f_{RY\phi}-6H^2\dot{\phi}(1-2q)(\dot{\phi}f_{Y\phi\phi}\\
&&+\dot{Y}f_{YY})-6H^2\dot{Y}(1-2q)(\dot{\phi}f_{YY\phi}+\dot{Y}f_{YYY})\geq0,
\label{15}\\
\nonumber&&\textbf{DEC:}~~\kappa^2(\rho-p)+f-R f_{R}-\left(5H\dot{R}
+\ddot{R}\right)f_{RR}-\left\{5H\dot{Y}+\ddot{Y}-2H^2\times\right.\\
\nonumber&&\left.(1-2q)\ddot{R}-\dot{R}H^3(14-16q)\right\}f_{RY}-\left(5H
\dot{\phi}+\ddot{\phi}\right)f_{R\phi}-\dot{R}^2f_{RRR}-2\dot{R}\times\\
\nonumber&&\left(\dot{Y}-\dot{R}H^2(1-2q)\right)f_{RRY}-2\dot{\phi}\dot{R}
f_{RR\phi}+\left(4H^2(1-2q)\dot{R}-\dot{Y}\right)\dot{Y}f_{RYY}\\
\nonumber&&-\dot{\phi}^2f_{R\phi\phi}+2\dot{\phi}\left(2H^2\dot{R}(1-2q)
-\dot{Y}\right)f_{RY\phi}+2H^2\dot{Y}\left(1-2q\right)\left(\dot{Y}f_{YYY}
\right.\\
\nonumber&&\left.+\dot{\phi}f_{YY\phi}\right)+2H^2\dot{\phi}\left(1-2q\right)
\left(\dot{Y}f_{YY\phi}+\dot{\phi}f_{Y\phi\phi}\right)+H^2\left\{2(1-2q)
\ddot{Y}+H\dot{Y}\right.\\
\nonumber&&\left.\times(14-16q)\right\}f_{YY}+H^2\left\{2(1-2q)\ddot{\phi}+H
\dot{\phi}(14-16q)\right\}+4H^4(s+j\\
&&-5q^2-q-5)f_{Y}\geq0.\label{19}
\end{eqnarray}

\section{Reconstruction of $f(R,Y,\phi)$ gravity}

In this section, we are presenting the reconstruction of
$f(R,Y,\phi)$ gravity by using well-known cosmological solutions
namely de-Sitter (dS) and power law cosmologies.

\subsection{de-Sitter Universe Models}

The dS solutions are very importance in cosmology to explain the
current cosmic epoch. The dS model is described by the exponential
scale factor, Hubble parameter  and Ricci tensor as,
\begin{equation}\label{26}
a(t) = a_{0}e^{H_{0}t},~~H = H_{0},~~R = 12H_{0}^{2}.
\end{equation}
In this reconstruction, we consider the matter source with constant EoS
parameter $w=\frac{p}{\rho}$ so that
\begin{equation}\label{27}
\rho=\rho_{0}e^{-3(1+w)H_{0}t},~~w\neq-1.
\end{equation}
Here we are using \cite{32}
\begin{equation}\label{35}
\omega(\phi)=\omega_{0}\phi^{m},~~\phi(t)\sim a(t)^{\beta}.
\end{equation}
Using these quantities along with Eqs.(\ref{26}) and (\ref{27}) in
Eq.(\ref{2}), we obtain
\begin{eqnarray}\label{36}
\nonumber\noindent&& 3 H_{0}^{2}\beta\phi f_{R\phi}-18 H_{0}^{4}\beta\phi
f_{Y\phi}-3H_{0}^{2}f_{R}+42H_{0}^{4}f_{Y}-\frac{1}{2}f(R,Y,\phi)+\frac{1}{2}
\beta^{2}\omega_{0}\\
&&\times H_{0}^{2}\phi^{m+2}-\kappa^2\rho_{0}a_{0}^{3(1+w)}
\phi^{-\frac{3}{\beta}}=0.
\end{eqnarray}
This is a second order partial differential equation which can be
converted in canonical form whose solution yields
\begin{equation}\label{28}
f(R,Y,\phi)=\alpha_{1}\alpha_{2}\alpha_{3}e^{\alpha_{1}R}e^{\alpha_{2}Y}
\phi^{\gamma_{1}}+\gamma_{2}\phi^{\gamma_{3}}+\gamma_{4}\phi^{\gamma_{5}},
\end{equation}
where $\alpha_{i}'s$ are constants of integration and
\begin{eqnarray}\label{36a}
\nonumber\gamma_{1}&=&\frac{18\beta\alpha _1 H_0^{2}-108\beta\alpha _2 H_{0}^4
-5+6\alpha _1H_{0}^2-84\alpha _2H_{0}^4}{6\left(H_{0}^2\alpha_1\beta-6\beta
\alpha_2H_{0}^4\right)}\\
\gamma_{2}&=&\omega_{0}\beta^{2} H_{0}^{2},~~\gamma_{3}=m+2,~~\gamma_{4}=
-2\kappa^2\rho_{0} a_{0}^{3(1+w)},~~\gamma_{5}=-\frac{3}{\beta}.
\end{eqnarray}
Introducing model (\ref{28}) in the energy conditions (\ref{9})-(\ref{12}) it
follows,
\begin{eqnarray}
\nonumber\noindent&&\textbf{NEC:}~~\kappa^2\rho_{0}e^{-3H_{0}(1+w)t}+\kappa^2 p-
\omega_{0}\beta^{2}H_{0}^{2}a_{0}^{\beta(m+2)}e^{\beta(m+2)H_{0}t}+\beta(\beta
\gamma_{1}-1)\alpha_{1}\alpha_{2}\\
\nonumber&&\times\alpha_{3}\gamma_{1}H_{0}^{2}\left(\alpha_{1}-6\alpha_{2}
H_{0}^{2}\right)a_{0}^{\beta\gamma_{1}}e^{\beta\gamma_{1}H_{0}t+12\alpha_{1}
H_{0}^{2}+36\alpha_{2}H_{0}^{4}}+\beta^{2}\alpha_{1}\alpha_{2}\alpha_{3}
\gamma_{1}\left(\gamma_{1}-1\right)\\
&&\times\left(\alpha_{1}-6\alpha_{2}H_{0}^{2}\right)H_{0}^{2}a_{0}^{\beta
\gamma_{1}}e^{\beta\gamma_{1}H_{0}t+12\alpha_{1}H_{0}^{2}+36\alpha_{2}
H_{0}^{4}}\geq0,\label{37a}\\
\nonumber&&\textbf{WEC:}~~\kappa^2\rho_{0}e^{-3H_{0}(1+w)t}+\frac{1}{2}\alpha_{1}
\alpha_{2}\alpha_{3}a_{0}^{\beta\gamma_{1}}e^{\beta\gamma_{1}H_{0}t
+12\alpha_{1}H_{0}^2+36\alpha_{2}H_{0}^4}\left(1-12\alpha_{1}H_{0}^2\right.\\
\nonumber&&\left.-36\alpha_{2}H_{0}^4\right)+\gamma_{2}a_{0}^{\beta\gamma_{3}}
e^{\beta\gamma_{3}H_{0}t}+\gamma_{4}a_{0}^{\beta\gamma_{5}}e^{\beta\gamma_{5}
H_{0}t}-\frac{1}{2}\omega_{0}\beta^{2}H_{0}^{2}a_{0}^{\beta(m+2)}e^{\beta(m+2)
H_{0}t}\\
&&-3\beta\alpha_{1}\alpha_{2}\alpha_{3}\gamma_{1}H_{0}^{2}\left(\alpha_{1}
-6\alpha_{2}H_{0}^{2}\right)a_{0}^{\beta\gamma_{1}}e^{\beta\gamma_{1}H_{0}t
+12\alpha_{1}H_{0}^{2}+36\alpha_{2}H_{0}^{4}}\geq0,\label{37b}\\
\nonumber&&\textbf{SEC:}~~\kappa^2\rho_{0}e^{-3H_{0}(1+w)t}+3\kappa^2 p+\alpha_{1}
\alpha_{2}\alpha_{3}a_{0}^{\beta\gamma_{1}}e^{\beta\gamma_{1}H_{0}t
+12\alpha_{1}H_{0}^2+36\alpha_{2}H_{0}^4}\left(36\alpha_{2}H_{0}^4\right.\\
\nonumber&&\left.+12\alpha_{1}H_{0}^2-1\right)-\gamma_{2}a_{0}^{\beta
\gamma_{3}}e^{\beta\gamma_{3}H_{0}t}-\gamma_{4}a_{0}^{\beta\gamma_{5}}e^{\beta
\gamma_{5}H_{0}t}-2\omega_{0}\beta^{2}H_{0}^{2}a_{0}^{\beta(m+2)}\times\\
\nonumber&&e^{\beta(m+2)H_{0}t}+3\beta(1+\beta)\alpha_{1}\alpha_{2}\alpha_{3}
\gamma_{1}H_{0}^{2}\left(\alpha_{1}-6\alpha_{2}H_{0}^{2}\right)e^{\beta
\gamma_{1}H_{0}t+12\alpha_{1}H_{0}^{2}+36\alpha_{2}H_{0}^{4}}\\
\nonumber&&\times a_{0}^{\beta\gamma_{1}}+3\beta^{2}\alpha_{1}\alpha_{2}
\alpha_{3}\gamma_{1}(\gamma_{1}-1)H_{0}^{2}\left(\alpha_{1}-6\alpha_{2}
H_{0}^{2}\right)a_{0}^{\beta\gamma_{1}}\\
&&e^{\beta\gamma_{1}H_{0}t+12\alpha_{1}H_{0}^{2}+36\alpha_{2}H_{0}^{4}}\geq0,
\label{37c}\\
\nonumber&&\textbf{DEC:}~~\kappa^2\rho_{0}e^{-3H_{0}(1+w)t}-\kappa^2 p+\alpha_{1}
\alpha_{2}\alpha_{3}a_{0}^{\beta\gamma_{1}}e^{\beta\gamma_{1}H_{0}t+12
\alpha_{1}H_{0}^2+36\alpha_{2}H_{0}^4}\left(1-12\alpha_{1}\right.\\
\nonumber&&\left.\times H_{0}^2-36\alpha_{2}H_{0}^4\right)+\gamma_{2}
a_{0}^{\beta\gamma_{3}}e^{\beta\gamma_{3}H_{0}t}+\gamma_{4}a_{0}^{\beta
\gamma_{5}}e^{\beta\gamma_{5}H_{0}t}+\beta(\beta+5)\alpha_{1}\alpha_{2}
\alpha_{3}\gamma_{1}H_{0}^{2}\\
\nonumber&&\times\left(6\alpha_{2}H_{0}^{2}-\alpha_{1}\right)a_{0}^{\beta
\gamma_{1}}e^{\beta\gamma_{1}H_{0}t+12\alpha_{1}H_{0}^{2}+36\alpha_{2}
H_{0}^{4}}+\beta^{2}\alpha_{1}\alpha_{2}\alpha_{3}\gamma_{1}(\gamma_{1}-1)
H_{0}^{2}\times\\
&&\left(6\alpha_{2}H_{0}^{2}-\alpha_{1}\right)a_{0}^{\beta\gamma_{1}}e^{\beta
\gamma_{1}H_{0}t+12\alpha_{1}H_{0}^{2}+36\alpha_{2}H_{0}^{4}}\geq0.\label{37d}
\end{eqnarray}

\begin{landscape}
\centering
\begin{table}[htbps]
\centering
\begin{tabular}{|l|c|c|} \hline \hline
Variations of $\alpha_{i}'s$
& Validity of WEC                                   & Validity of NEC \\ \hline \hline
& $\alpha_{3}=0$, $\forall$ $\beta$ \& $m$          & $\alpha_{3}<0$ with $\forall$ $m$ \& $\beta$ \\ \cline{2-2} \cline{3-2}
$\alpha_{1}>0$,$\alpha_{2}>0$
& $\alpha_{3}>0$, $\beta\geq 0$, $\forall$ $m$      & $\alpha_{3}=0$ with ($m\geq0$, $\beta\leq-1$) or ($m\leq-2.8$, $\beta \geq 2$) \\ \cline{2-2}
& $\alpha_{3}<0$, $\beta\leq-1$, $\forall$ $m$      & \\ \hline
$\alpha_{1}<0$, $\alpha_{2}>0$
&  $\forall$ $\alpha_{3}$, $m$ \& $\beta\geq0$    & $\alpha_{3}=0$ with ($m\geq0$, $\beta\leq-2$) or ($m\leq-2.8$, $\beta \geq 2$) \\ \cline{3-2}
&                                                 & $\alpha_{3}>0$ with $\beta>0$ or $\beta<0$ $\forall$ $m$\\ \hline
$\alpha_{1}>0$, $\alpha_{2}<0$
& $\forall$ $\alpha_{3}$, $\beta$, $m$ \& $t>3.6$  & $\forall$ $\alpha_{3}$ with ($m\geq0$, $\beta \leq -1.5$) or ($m\leq-3$, $\beta\geq2.8$) \\ \hline
$\alpha_{1}<0$, $\alpha_{2}<0$
& $\forall$ $\alpha_{3}$, $\beta$, $m$ \& $t\geq3.6$  & $\forall$ $\alpha_{3}$ with ($m\geq0$,  $\beta\leq-1.4$) or ($m\leq-3.6$,
$\beta\geq1$) \\ \hline \hline

& $\alpha_{1}<0$ with $\beta\leq-1$, $\forall$ $m$    & $\alpha_{1}<0$ with $\beta>0$ or $\beta<0$ \& $\forall$ $m$ \\ \cline{2-3}\cline{3-2}
$\alpha_{2}>0$, $\alpha_{3}>0$
& $\alpha_{1}=0$ with $t\geq3.6$, $\forall$ $m$ \& $\beta$    & $\alpha_{1}=0$ with ($m\leq-3$, $\beta \geq 2.8$) or
($m\geq0$, $\beta\leq-1$)  \\ \cline{2-2}
& $\alpha_{1}>0$ with $\beta>0$, $t\geq3.6$, $\forall$ $m$     &  \\ \hline

& $\alpha_{1}<0$, $\beta>0$, $\forall$ $m$        & $\alpha_{1}=0$ with ($m\leq-3$, $\beta\geq1$) or ($m\geq-1$, $\beta\leq-1$)  \\
\cline{2-3}\cline{3-2}
$\alpha_{2}>0$, $\alpha_{3}<0$
& $\alpha_{1}>0$, $\beta\leq-1$, $\forall$ $m$               & $\alpha_{1}>0$ with $\beta<0$ or $\beta>0$ \& $\forall$ $m$  \\ \cline{2-2}
& $\alpha_{1}=0$, $t\geq3.6$, $\forall$ $\beta$ \& $m$       &  \\ \hline
$\alpha_{2}<0$, $\alpha_{3}>0$
& $\forall$ $\alpha_{1}$, $\beta$ \& $m$ with $t\geq3.6$     & $\forall$ $\alpha_{1}$ with ($\beta\leq-1$, $m\geq1$) or
($\beta\geq2.8$, $m\leq-3$) \\ \hline
$\alpha_{2}<0$, $\alpha_{3}<0$
& $\forall$ $\alpha_{1}$, $\beta$ \& $m$ with $t\geq3.6$  & $\forall$ $\alpha_{1}$ with ($\beta\leq-1$, $m\geq0.8$) or
($\beta\geq2.5$, $m\leq-3.5$) \\ \hline \hline

$\alpha_{1}>0$, $\alpha_{3}>0$
& $\alpha_{2}>0$ with $\beta>0$, $\forall$ $m$            & $\alpha_{2}\leq0$ with ($\beta\leq-1.5$, $m\geq0$) or
($\beta\geq2.8$, $m\leq-3$) \\
\cline{2-2}
& $\alpha_{2}\leq0$ with $\forall$ $\beta$, $m$ \& $t\geq3.6$       &    \\ \hline
$\alpha_{1}>0$, $\alpha_{3}<0$
& $\alpha_{2}\leq0$ with $\forall$ $\beta$, $m$ \& $t\geq3.6$       & $\forall$ $\alpha_{2}$ with ($\beta\leq-2$, $m\geq0$) or
($\beta\geq1$, $m\leq-3$) \\
\cline{2-2}
& $\alpha_{2}>0$ with $\beta\leq-0.5$ \& $\forall$ $m$        &   \\ \hline
$\alpha_{1}<0$, $\alpha_{3}>0$
& $\alpha_{2}\leq0$ with $\forall$ $\beta$, $m$ \& $t\geq3.6$       & $\alpha_{2}>0$ with $\forall$ $\beta$ \& $m$ \\
\cline{2-2}\cline{3-2}
&  $\alpha_{2}>0$ with $\beta\leq-0.5$ \& $\forall$ $m$       & $\alpha_{2}\leq0$ with ($\beta\geq2.8$, $m\leq-3$) or
($\beta\leq-1.4$, $m\geq0$) \\ \hline
$\alpha_{1}<0$, $\alpha_{3}<0$
& $\alpha_{2}>0$, $\beta\geq0$ \& $m$    & $\alpha_{2}\leq0$ with ($\beta\geq2$, $m\leq-3.5$) or ($\beta\leq-1.4$, $m\geq0$)  \\
\cline{2-2}
& $\alpha_{2}\leq0$ with $\forall$ $\beta$, $m$ \& $t\geq3.6$       &  \\ \hline \hline
\end{tabular}
\caption{Validity regions of WEC and NEC for dS $f(R,Y,\phi)$ model.}
\label{Table1}
\end{table}
\end{landscape}

The inequalities (\ref{37a})-(\ref{37d}) depend on six parameters
$\alpha_{1}$, $\alpha_{2}$, $\alpha_{3}$, $\beta$, $m$ and $t$. In this
approach, we fix two parameters and find the viable region by exploring the
possible ranges of other parameters. We prefer to fix integration constants
and show the results for WEC and NEC. Herein, we set the present day values
of Hubble parameter, fractional energy density and cosmographic parameters as
$H_{0}=67.3$, $\Omega_{m0}= 0.315$ \cite{32a} $q=-0.81$, $j=2.16$, $s=-0.22$,
\cite{13}. The viability regions for all the possible cases for dS
$f(R,Y,\phi)$ model are presented in Table \ref{Table1}.

Initially, we vary $\alpha_{1}$ and $\alpha_{2}$ to check the validity of WEC
and NEC for different values of $\alpha_{3}$, $\beta$ and $m$. If we set both
$\alpha_{1}$ and $\alpha_{2}$ as positive then WEC is valid for $m$, however
$\beta$ needs some particular ranges as: ($\alpha_{3}>0$, $\beta\geq0$),
($\alpha_{3}=0$, $\forall$ $\beta$) and ($\alpha_{3}<0$, $\beta\leq-1$). NEC
is valid only if $\alpha_3\leq0$ and the suitable regions are ($\alpha_3<0$,
$\forall$ $m$, $\beta$), ($\alpha_3=0$, $m\geq0$, $\beta\leq-1$) and
($\alpha_3=0$, $m\leq-2.8$, $\beta\geq2$). In Fig.\textbf{1}, we present the
evolution of WEC and NEC to show some viable regions in this case.
\begin{figure}
\centering \epsfig{file=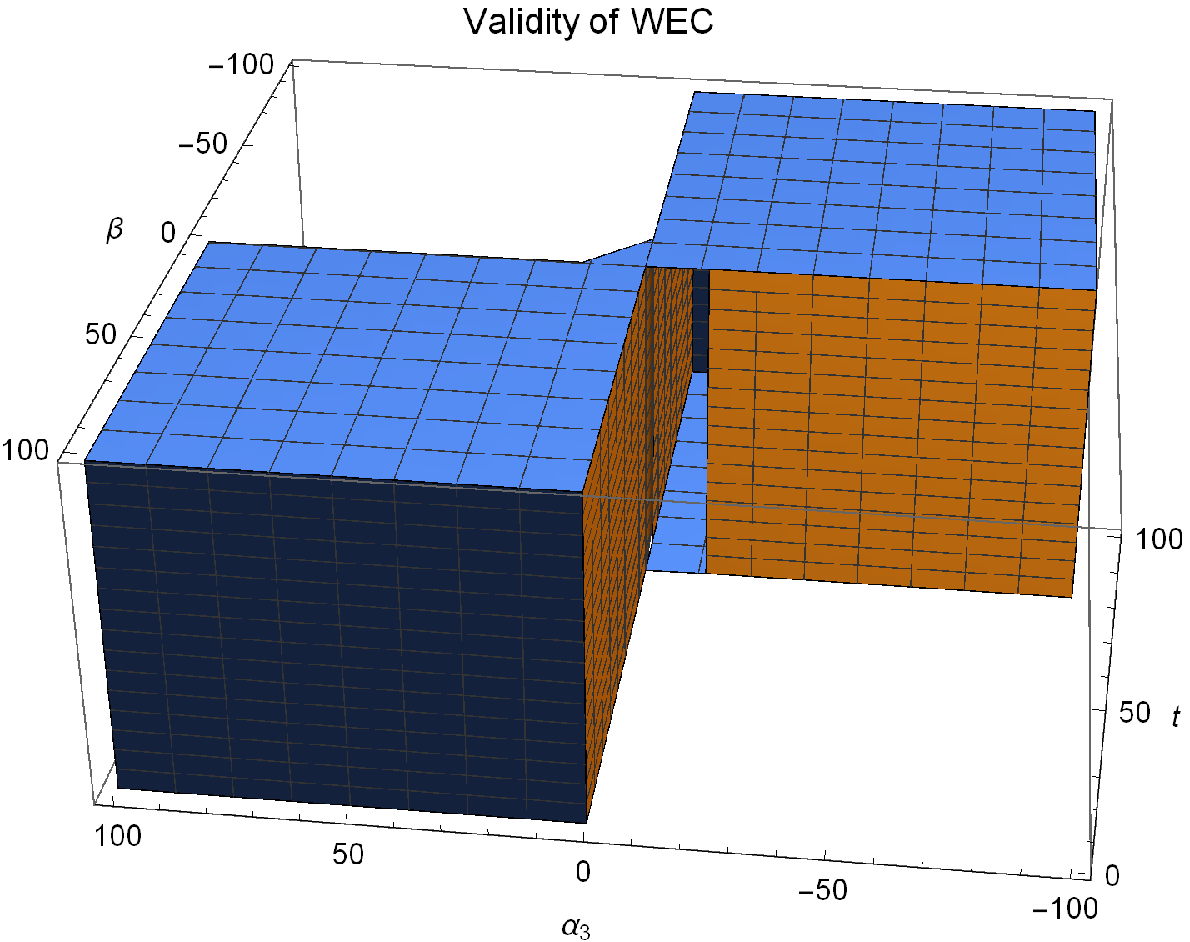, width=.45\linewidth,
height=2in}\epsfig{file=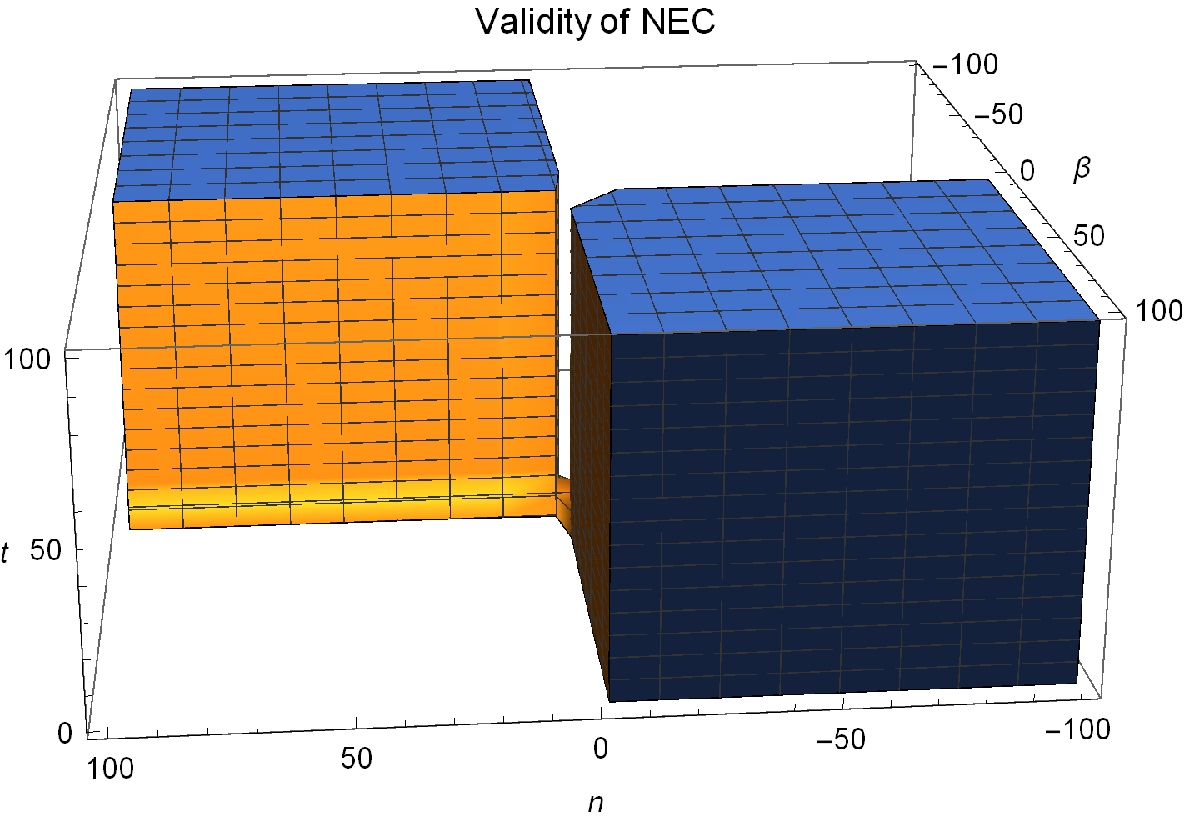, width=.45\linewidth,
height=2in}
\caption{Variation of energy constraints for dS $f(R,Y,\phi)$ model with
$\alpha_1>0$ and $\alpha_2>0$. In left plot we set $m=-10$ (one can set
any value since results are valid for all $m$) and show the variation for all
$\alpha_3$ and $\beta$. Right plot shows the validity regions of NEC for
$\alpha_3=0$.}
\end{figure}
If $\alpha_{1}<0$ and $\alpha_{2}>0$, WEC is valid for all values of
$\alpha_{3}$ $\&$ $m$ with $\beta\geq0$. For $\alpha_{3}>0$, NEC is valid for
all values of $m$ and $\beta$ except $\beta=0$ and if $\alpha_3=0$ then the
validity of NEC requires ($m\geq0$, $\beta\leq-2$) or ($m\leq-2.8$, $\beta
\geq 2$). If $\alpha_{1}>0$ and $\alpha_{2}<0$, WEC is valid for all values
of $\alpha_{3}$, $\beta$ and $m$ with $t>3.6$, in case of NEC we require
($\beta \leq -1.5$, $m\geq0$), ($\beta\geq2.8$, $m\leq-3$) for all
$\alpha_3$. For choosing $\alpha_{1}<0$ and $\alpha_{2}<0$, WEC is valid for
all values of $\alpha_{3}$, $\beta$ and $m$ with $t\geq3.6$. For all values
of $\alpha_{3}$, NEC is valid for $\beta\leq-1.4$ with $m\geq0$ and for
$\beta\geq1$ with $m\leq-3.6$.

Now we are varying $\alpha_{2}$ and $\alpha_{3}$, starting with
$\alpha_{2}>0$ and $\alpha_{3}>0$. For $\alpha_{1}>0$, WEC is valid for all
values of $m$ with $\beta>0$ $\&$ $t\geq3.6$ and NEC violates and for
$\alpha_{1}<0$ WEC is valid for all values of $m$ with $\beta\leq-1$ and NEC
is valid for all values of $m$ $\&$ $\beta$ except $\beta=0$. For
$\alpha_{1}=0$, WEC is valid for all values of $m$ with $\beta\leq-1$ and NEC
is valid for $m\geq0$ with $\beta\leq-1$ and for $m\leq-3$ with $\beta \geq
2.8$. In case of $\alpha_{2}>0$ and $\alpha_{3}<0$, the validity of WEC and
NEC establishes three cases: $(i)$ if $\alpha_{1}<0$, WEC is valid for all
values of $m$ with $\beta>0$ and NEC violate, $(ii)$ if $\alpha_{1}>0$, WEC
is valid for all values of $m$ with $\beta\leq-1$ and NEC is valid for all
values of $m$ $\&$ $\beta$ except $\beta=0$, $(iii)$ if $\alpha_{1}=0$, WEC
is valid for all values of $\beta$ $\&$ $m$ with $t\geq3.6$ and NEC is valid
for $\beta\geq1$ with $m\leq-3$ and for $\beta\leq-1$ with $m\geq-1$. For
$\alpha_{2}<0$ and $\alpha_{3}>0$, WEC is satisfied for all values of
$\alpha_{1}$, $\beta$ $\&$ $m$ with $t\geq3.6$ whereas the validity of NEC
requires ($\beta\leq-1$, $m\geq1$) or ($\beta\geq2.8$, $m\leq-3$) for all
$\alpha_1$. Similarly, for $\alpha_{2}<0$ and $\alpha_{3}<0$, WEC is valid
for all values of $\alpha_{1}$, $\beta$, and $m$ with $t\geq3.6$ whereas the
validity of NEC requires ($\beta\leq-1$, $m\geq0.8$) or ($\beta\geq2.5$,
$m\leq-3.5$) for all $\alpha_1$.

Next we are varying $\alpha_{1}$ and $\alpha_{3}$, taking $\alpha_{1}$ and
$\alpha_{3}$ both as positive. For $\alpha_{2}>0$, WEC is valid for all
values of $m$ with $\beta>0$ and NEC violates. For $\alpha_{2}\leq0$, WEC is
valid for all values of $\beta$ $\&$ $m$ with $t\geq3.6$ and NEC is valid
for $\beta\leq-1.5$ with $m\geq0$ and for $\beta\geq2.8$ with $m\leq-3$. Now
taking $\alpha_{1}$ as positive and $\alpha_{3}$ as negative. For
$\alpha_{2}>0$, WEC is valid for all values of $m$ with $\beta\leq-0.5$ and
for $\alpha_{2}\leq0$ WEC is valid for all values of $\beta$ $\&$ $m$ with
$t\geq3.6$. For all values of $\alpha_{2}$ NEC is valid for $\beta\leq-2$
with $m\geq0$ and for $\beta\geq1$ with $m\leq-3$. Taking $\alpha_{1}$ as
negative and $\alpha_{3}$ as positive. For $\alpha_{2}>0$, NEC is valid for
all values of $m$ with $\beta\leq-0.5$ and WEC is valid for all values of
$\beta$ $\&$ $m$. For $\alpha_{2}\leq0$, WEC is valid for all values of
$\beta$ $\&$ $m$ with $t\geq3.6$ and NEC is valid for $\beta\geq2.8$ with
$m\leq-3$ and for $\beta\leq-1.4$ with $m\geq0$. Taking $\alpha_{1}$ and
$\alpha_{3}$ both as negative. For $\alpha_{2}>0$, WEC is valid for all
values of $m$ with $\beta\geq0$ and NEC violates. For $\alpha_{2}\leq0$ WEC
is valid for all values of $\beta$ $\&$ $m$ with $t\geq3.6$ and NEC is valid
for $\beta\geq2$ with $m\leq-3.5$ and for $\beta\leq-1.4$ with $m\geq0$.

\begin{itemize}
\item \textbf{de-Sitter model independent of Y}
\end{itemize}
Here we are taking function $f(R,\phi)$ and inserting Eq.(\ref{35}) along with
Eqs.(\ref{26}) and (\ref{27}) in Eq. (\ref{2}) we obtain
\begin{equation}\label{37e}
3H_{0}^{2}\beta\phi f_{R\phi}-3H_{0}^{2}f_{R}-\frac{1}{2}f(R,\phi)+\frac{1}{2}
\omega_{0}\beta^{2}H_{0}^{2}\phi^{m+2}-\kappa^2\rho_{0}a_{0}^{3(1+w)}
\phi^{-\frac{3}{\beta}}=0.
\end{equation}
Solving this equation we have,
\begin{equation}\label{29}
f(R,\phi)=\alpha_{1}\alpha_{2}e^{\alpha_{1}R}\phi^{\gamma_{1}}+\gamma_{2}
\phi^{\gamma_{3}}+\gamma_{4}\phi^{\gamma_{5}},
\end{equation}
where $\alpha_{i}'s$ are constants of integration and
\begin{eqnarray}\label{37f}
\nonumber\gamma_{1}&=&-\frac{1}{\beta}(1+\frac{1}{6H_{0}^{2}\alpha_{1}}),
~~\gamma_{2}=\omega_{0}\beta^{2} H_{0}^{2},\\
\gamma_{3}&=&m+2,~~\gamma_{4}=-2 \kappa^2 \rho_{0}a_{0}^{3(1+w)},~~\gamma_{5}=
-\frac{3}{\beta}.
\end{eqnarray}
Introducing model (\ref{29}) in inequalities (\ref{9})-(\ref{12}) it follows,
\begin{eqnarray}
\nonumber&&\textbf{NEC:}~~\kappa^2\rho_{0}e^{-3H_{0}(1+w)t}+\kappa^2 p-\omega_{0}
\beta^{2}H_{0}^{2}a_{0}^{\beta(m+2)}e^{\beta(m+2)H_{0}t}+\beta(\beta-1)\times\\
\nonumber&&\alpha_{1}^{2}\alpha_{2}\gamma_{1}H_{0}^{2}a_{0}^{\beta\gamma_{1}}
e^{\beta\gamma_{1}H_{0}t+12\alpha_{1}H_{0}^{2}}+\beta^{2}\alpha_{1}^{2}
\alpha_{2}\gamma_{1}\left(\gamma_{1}-1\right)H_{0}^{2}a_{0}^{\beta\gamma_{1}}
\times\\
&&e^{\beta\gamma_{1}H_{0}t+12\alpha_{1}H_{0}^{2}}\geq0,\label{38a}\\
\nonumber&&\textbf{WEC:}~~\kappa^2\rho_{0}e^{-3H_{0}(1+w)t}+\frac{1}{2}\alpha_{1}
\alpha_{2}a_{0}^{\beta\gamma_{1}}e^{\beta\gamma_{1}H_{0}t+18\alpha_{1}H_{0}^{2}}
+\frac{1}{2}\gamma_{4}a_{0}^{\beta\gamma_{5}}e^{\beta\gamma_{5}
H_{0}t}\\
&&-6\alpha_{1}^{2}\alpha_{2}H_{0}^{2}a_{0}^{\beta\gamma_{1}}
e^{\beta\gamma_{1}H_{0}t+12\alpha_{1}H_{0}^{2}}-3\beta\alpha_{1}^{2}\alpha_{2}
\gamma_{1}H_{0}^{2}a_{0}^{\beta\gamma_{1}}e^{\beta\gamma_{1}H_{0}t+12\alpha_{1}
H_{0}^{2}}\geq0,\label{38b}\\
\nonumber&&\textbf{SEC:}~~\kappa^2\rho_{0}e^{-3H_{0}(1+w)t}+3\kappa^2 p-\alpha_{1}
\alpha_{2}a_{0}^{\beta\gamma_{1}}e^{\beta\gamma_{1}H_{0}t+12\alpha_{1}
H_{0}^{2}}-\gamma_{2}a_{0}^{\beta\gamma_{3}}e^{\beta\gamma_{3}H_{0}t}\\
\nonumber&&-\gamma_{4}a_{0}^{\beta\gamma_{5}}e^{\beta\gamma_{5}H_{0}t}
-2\omega_{0}\beta^{2}H_{0}^{2}a_{0}^{\beta(m+2)}e^{\beta(m+2)H_{0}t}+12
\alpha_{1}^{2}\alpha_{2}H_{0}^{2}a_{0}^{\beta\gamma_{1}}\times\\
\nonumber&&e^{\beta\gamma_{1}H_{0}t+12\alpha_{1}H_{0}^{2}}+3\beta(1+\beta)
\alpha_{1}^{2}\alpha_{2}\gamma_{1}H_{0}^{2}a_{0}^{\beta\gamma_{1}}e^{\beta
\gamma_{1}H_{0}t+12\alpha_{1}H_{0}^{2}}+3\beta^{2}\alpha_{1}^{2}\alpha_{2}
\gamma_{1}\\
&&\times(\gamma_{1}-1)H_{0}^{2}a_{0}^{\beta\gamma_{1}}e^{\beta\gamma_{1}H_{0}t
+12\alpha_{1}H_{0}^{2}}\geq0,\label{38c}\\
\nonumber&&\textbf{DEC:}~~\kappa^2\rho_{0}e^{-3H_{0}(1+w)t}-\kappa^2 p+\alpha_{1}
\alpha_{2}a_{0}^{\beta\gamma_{1}}e^{\beta\gamma_{1}H_{0}t+12\alpha_{1}H_{0}^{2}}
+\gamma_{2}a_{0}^{\beta\gamma_{3}}e^{\beta\gamma_{3}H_{0}t}\\
\nonumber&&+\gamma_{4}a_{0}^{\beta\gamma_{5}}e^{\beta\gamma_{5}H_{0}t}
-12\alpha_{1}^{2}\alpha_{2}H_{0}^{2}a_{0}^{\beta\gamma_{1}}e^{\beta\gamma_{1}
H_{0}t+12\alpha_{1}H_{0}^{2}}-\beta(\beta+5)\alpha_{1}^{2}\alpha_{2}\gamma_{1}
\times\\
&& H_{0}^{2}a_{0}^{\beta\gamma_{1}}e^{\beta\gamma_{1}H_{0}t+12
\alpha_{1}H_{0}^{2}}-\beta^{2}\alpha_{1}^{2}\alpha_{2}\gamma_{1}(\gamma_{1}-1)
H_{0}^{2}a_{0}^{\beta\gamma_{1}}e^{\beta\gamma_{1}H_{0}t+12\alpha_{1}H_{0}^{2}}
\geq0.\label{38d}
\end{eqnarray}

Here, we discuss the energy constraints for dS $f(R,\phi)$ model, the
inequalities representing these conditions depend on five parameters namely,
$\alpha_{1}$, $\alpha_{2}$, $\beta$, $m$ and $t$. One can see that WEC only
depends on $\alpha_{1}$, $\alpha_{2}$ and $t$. We find that WEC is satisfied
for two cases depending on the choice of $\alpha_1$: $(i)$ $\alpha_1>0$ with
$\alpha_2\geq0$ $(ii)$ $\alpha_1<0$ with for all $\alpha_2$. Now we discuss
NEC for three viable cases depending on the choice of $\alpha_{1}$ and
$\alpha_{2}$. If both $\alpha_{1}$ and $\alpha_{2}$ are positive then NEC is
valid for ($\beta<0$ with $m>-2$) and ($\beta>0$ with $m\leq-2$). Taking
$\alpha_{1}$ as negative and $\alpha_{2}$ as positive, NEC is valid for
$\beta\geq3$ with $m\leq-5$ and for $\beta\leq-1$ with $m\geq0.8$, similarly
for $\alpha_{1}<0$, $\alpha_{2}<0$ the validity of NEC requires
$\beta\geq3.5$ with $m\leq-5$ and $\beta\leq-1$ with $m\geq1$.
\begin{itemize}
\item \textbf{de-Sitter model independent of R}
\end{itemize}
Now we are taking function $f(Y,\phi)$ and inserting Eq.(\ref{35}) along with
Eqs.(\ref{26}) and (\ref{27}) in Eq.(\ref{2}) we get
\begin{equation}\label{38e}
18H_{0}^{4}\beta\phi f_{Y\phi}-42H_{0}^{4}f_{Y}+\frac{1}{2}f(Y,\phi)
-\frac{1}{2}\omega_{0}\beta^{2}H_{0}^{2}\phi^{m+2}-\kappa^2\rho_{0}a_{0}^{3(1+w)}
\phi^{-\frac{3}{\beta}}=0.
\end{equation}
whose solution yields
\begin{equation}\label{30}
f(Y,\phi)=\alpha_{1}\alpha_{2}e^{\alpha_{1}Y}\phi^{\gamma_{1}}+\gamma_{2}
\phi^{\gamma_{3}}+\gamma_{4}\phi^{\gamma_{5}},
\end{equation}
where $\alpha_{i}'s$ are constants of integration and
\begin{eqnarray}
\nonumber\gamma_{1}&=&-\frac{7}{3\beta}+\frac{1}{36H_{0}^{4}\alpha_{1}\beta},
~~\gamma_{2}=\omega_{0}\beta^{2} H_{0}^{2},\\
\gamma_{3}&=& m+2,~~\gamma_{4}=-2\kappa^2\rho_{0}a_{0}^{3(1+w)},~~\gamma_{5}=
-\frac{3}{\beta}.\label{38f}
\end{eqnarray}
Using model (\ref{30}) in constraints (\ref{9})-(\ref{12}) it follows,
\begin{eqnarray}
\nonumber&&\textbf{NEC:}~~\kappa^2\rho_{0}e^{-3H_{0}(1+w)t}+\kappa^2 p-\omega_{0}
\beta^{2}H_{0}^{2}a_{0}^{\beta(m+2)}e^{\beta(m+2)H_{0}t}-6\beta(\beta-1)\\
\nonumber&&\times\alpha_{1}^{2}\alpha_{2}\gamma_{1}H_{0}^{4}a_{0}^{\beta
\gamma_{1}}e^{\beta\gamma_{1}H_{0}t+36\alpha_{1}H_{0}^{4}}-6\beta^{2}
\alpha_{1}^{2}\alpha_{2}\gamma_{1}\left(\gamma_{1}-1\right)H_{0}^{4}
a_{0}^{\beta\gamma_{1}}\\
&&e^{\beta\gamma_{1}H_{0}t+36\alpha_{1}H_{0}^{4}}\geq0,\label{39a}\\
\nonumber&&\textbf{WEC:}~~\kappa^2\rho_{0}e^{-3H_{0}(1+w)t}+\frac{1}{2}\alpha_{1}
\alpha_{2}a_{0}^{\beta\gamma_{1}}e^{\beta\gamma_{1}H_{0}t+36\alpha_{1}H_{0}^{4}}
+\frac{1}{2}\gamma_{4}a_{0}^{\beta\gamma_{5}}e^{\beta\gamma_{5}
H_{0}t}\\
&&+18\alpha_{1}^{2}\alpha_{2}\gamma_{1}H_{0}^{4}a_{0}^{\beta\gamma_{1}}
e^{\beta\gamma_{1}H_{0}t+36\alpha_{1}H_{0}^{4}}-18\alpha_{1}^{2}\alpha_{2}
H_{0}^{4}a_{0}^{\beta\gamma_{1}}e^{\beta\gamma_{1}H_{0}t+36\alpha_{1}H_{0}^{4}}
\geq0,\label{39b}\\
\nonumber&&\textbf{SEC:}~~\kappa^2\rho_{0}e^{-3H_{0}(1+w)t}+3 \kappa^2 p -\alpha_{1}
\alpha_{2} a_{0}^{\beta \gamma_{1}} e^{\beta \gamma_{1} H_{0}t + 36\alpha_{1}
H_{0}^{4}}-\gamma_{2}a_{0}^{\beta\gamma_{3}}e^{\beta\gamma_{3}H_{0}t}\\
\nonumber&&-\gamma_{4}a_{0}^{\beta\gamma_{5}}e^{\beta\gamma_{5}H_{0}t}
-2\omega_{0}\beta^{2}H_{0}^{2}a_{0}^{\beta(m+2)}e^{\beta(m+2)H_{0}t}-18\beta
(1+\beta)\alpha_{1}^{2}\alpha_{2}\gamma_{1}H_{0}^{4}a_{0}^{\beta\gamma_{1}}\\
\nonumber&&\times e^{\beta\gamma_{1}H_{0}t+36\alpha_{1}H_{0}^{4}}-18\beta^{2}
\alpha_{1}^{2} \alpha_{2} \gamma_{1} (\gamma_{1}-1) H_{0}^{4} a_{0}^{\beta
\gamma_{1}} e^{\beta\gamma_{1} H_{0}t +36\alpha_{1}H_{0}^{4}}+36\alpha_{1}^{2}
\alpha_{2}\\
&&\times H_{0}^{4}a_{0}^{\beta\gamma_{1}}e^{\beta\gamma_{1}H_{0}t+36\alpha_{1}
H_{0}^{4}}\geq0,\label{39c}\\
\nonumber&&\textbf{DEC:}~~\kappa^2 \rho_{0} e^{-3H_{0}(1+w)t}-\kappa^2 p +\alpha_{1}
\alpha_{2}a_{0}^{\beta\gamma_{1}} e^{\beta\gamma_{1}H_{0}t+36\alpha_{1}
H_{0}^{4}}+\gamma_{2}a_{0}^{\beta\gamma_{3}}e^{\beta\gamma_{3}H_{0}t}\\
\nonumber&&+\gamma_{4}a_{0}^{\beta\gamma_{5}}e^{\beta\gamma_{5}H_{0}t}+6\beta
(\beta+5)\alpha_{1}^{2}\alpha_{2}\gamma_{1}H_{0}^{4}a_{0}^{\beta\gamma_{1}}
e^{\beta\gamma_{1}H_{0}t+36\alpha_{1}H_{0}^{4}}+6\beta^{2}\alpha_{1}^{2}
\alpha_{2}\gamma_{1}\times\\
&& (\gamma_{1}-1)H_{0}^{4}a_{0}^{\beta\gamma_{1}}e^{\beta\gamma_{1}H_{0}t+36
\alpha_{1}H_{0}^{4}}-36\alpha_{1}^{2}\alpha_{2}H_{0}^{4}a_{0}^{\beta\gamma_{1}}
e^{\beta\gamma_{1}H_{0}t+36\alpha_{1}H_{0}^{4}}\geq0.\label{39d}
\end{eqnarray}

Here, WEC depends only on $\alpha_{1}$, $\alpha_{2}$ and $t$ as in previous
case. We find that WEC is satisfied only if $\alpha_2\leq0$ for all values of
$\alpha_{1}$. Now we discuss the validity of NEC by varying $\alpha_{1}$ and
$\alpha_{2}$. If $\alpha_{1}$ and $\alpha_{2}$ both are positive then NEC
violates whereas for all other cases, ($\alpha_{1}<0$, $\alpha_{2}>0$),
($\alpha_{1}>0$, $\alpha_{2}<0$) and ($\alpha_{1}<0$, $\alpha_{2}<0$) it is
valid for all values of $m$ and $\beta$ except $\beta=0$.

\subsection{Power Law Solutions}

It would be very useful to discuss power solutions in this modified theory
according to different phases of cosmic evolution. These solutions are
helpful to explain all cosmic evolutions such as dark energy, matter and
radiation dominated eras. We are discussing power law solutions for two
models of $f(R,Y,\phi)$ gravity. The scale factor for this model is defined
as \cite{13,33}
\begin{equation}\label{25}
a(t)=a_{0}t^{n},~~H(t)=\frac{n}{t},~~R = 6n(1 - 2n)t^{-2},
\end{equation}
where $n>0$. For decelerated universe we have $0<n<1$, which leads to dust
dominated $(n=\frac{2}{3})$ or radiation dominated $(n=\frac{1}{2})$ while
$n>1$ leads to accelerating picture of the universe.

\begin{itemize}
\item \textbf{Power Law Solution independent of R}
\end{itemize}
Here, we are taking function $f(Y,\phi)$, inserting Eqs.(\ref{27}), (\ref{35})
and (\ref{25}) in Eq.(\ref{2}) we obtain
\begin{eqnarray}\label{25a}
\nonumber&&\frac{2(3n-2)}{4n^{2}-3n+1}Y^{2}f_{YY}-\frac{n(3n-2)}{2(4n^{2}
-3n+1)}\phi Yf_{Y\phi}+\frac{7n^{2}-19n+4}{2(4n^{2}-3n+1)}Y f_{Y}\\
&&-\frac{1}{2}f-\kappa^2\rho_{0}a_{0}^{3(1+w)}\phi^{-\frac{3}{\beta}}+\frac{1}{2}
\omega_{0}\beta^{2}n^{2}a_{0}^{\frac{2}{n}}\phi^{m+2-\frac{2}{n\beta}}=0,
\end{eqnarray}
whose solution results in following $f(Y,\phi)$ model
\begin{equation}\label{31}
f(Y,\phi)=\alpha_{1}\alpha_{2}\phi^{\gamma_{1}}Y^{\gamma_{2}}+\gamma_{3}
\phi^{\gamma_{4}}+\gamma_{5}\phi^{\gamma_{6}},
\end{equation}
where $\alpha_{i}'s$ are constants of integration and
\begin{eqnarray}\label{25b}
\nonumber\gamma_{1}&=&\frac{2(3n-2)\alpha_{1}}{4n^{2}-3n+1}+\frac{7n^{2}-31n
+12}{n(3n-2)}-\frac{2(4n^{2}-3n+1)^{2}}{n^{2}(3n-2)^{2}\alpha_{1}},\\
\nonumber\gamma_{2}&=&\frac{n(3n-2)\alpha_{1}}{2(4n^{2}-3n+1)},~~~\gamma_{3}=
-\omega_{0}\beta^{2}n^{2}a_{0}^{\frac{2}{n}},~~~\gamma_{4}=m+2-\frac{2}
{n\beta},\\
\gamma_{5}&=&-2 \kappa^2 \rho_{0}a_{0}^{3(1+w)},~~~\gamma_{6}=-\frac{3}{\beta}.
\end{eqnarray}
Introducing (\ref{31}) in the energy constraints (\ref{9})-(\ref{12}), one
can find the inequalities for this model depend on six parameters
$\alpha_{1}$, $\alpha_{2}$, $\beta$, $m$, $n$ and $t$. We will only discuss
the WEC and NEC for different values of $\beta$ and $m$ by fixing $n$ and
$\alpha_{i}$'s where $i=1,2$. Starting with $\alpha_{1}$ and $\alpha_{2}$
both as positive, WEC is valid for $n>1$ with $\beta\leq-0.1$, $m\geq0$,
$t\geq1.1$ and NEC is valid for all values of $m$ with $n>1$ $\&$
$\beta\geq0$. Now taking $\alpha_{1}$ as negative and $\alpha_{2}$ as
positive, WEC is valid for $1<n\leq1.8$ with $\beta\leq-3$, $m\geq0$ and for
$n\geq2.3$ with $\beta\geq2$ $\&$ $m\leq-1$. Similarly, NEC is valid for all
values of $m$ with $n>1$, $\beta\leq-0.12$ and $t\geq1.01$. Now taking
$\alpha_{1}$ as positive and $\alpha_{2}$ as negative, WEC is valid for
$n\geq1.7$ with $\beta\geq0.1$ $\&$ $m\leq-10$ and NEC is valid for all
values of $m$ with $n>1$, $\beta\geq0$ $\&$ $t\geq1.07$. Taking $\alpha_{1}$
and $\alpha_{2}$ both as negative, WEC is valid for $1<n\leq1.9$ with
$\beta>0$, $m\leq-6.5$ $\&$ $t>1$ and for $n\geq2$ WEC is valid for
$\beta\leq0$ with $m\geq4$. In this case NEC is valid for $1<n\leq1.5$ with
$\beta\geq0$, $m\leq-2.6$ $\&$ $t\geq1.9$ and for $n\geq2$ it is valid for
$\beta<0$ with $m\geq0$, $t\geq1.05$ and for $\beta\geq0$ with $m\leq-4$,
$t\geq1.08$.

\begin{itemize}
\item \textbf{Power Law Solution independent of Y}
\end{itemize}
Now we are taking function $f(R,\phi)$, inserting Eq.(\ref{27}) along with
Eqs.(\ref{35}) and (\ref{25}) in Eq.(\ref{2}) yields
\begin{eqnarray}\label{25c}
\nonumber&&\frac{1}{3n-1}R^{2}f_{RR}+\frac{n-1}{2(3n-1)}R f_{R}-\frac{n\beta}
{2(3n-1)}\phi R f_{R\phi}-\kappa^2\rho_{0}a_{0}^{3(1+w)}\phi^{-\frac{3}{\beta}}\\
&&-\frac{1}{2}f+\frac{1}{2}\omega_{0}\beta^{2}n^{2}a_{0}^{\frac{2}{n}}
\phi^{m+2-\frac{2}{n\beta}}=0.
\end{eqnarray}
Solving this we have,
\begin{equation}\label{32}
f(R,\phi)=\alpha_{1}\alpha_{2}\phi^{\gamma_{1}}R^{\gamma_{2}}+\gamma_{3}
\phi^{\gamma_{4}}+\gamma_{5}\phi^{\gamma_{6}},
\end{equation}
where $\alpha_{i}'s$ are constants of integration and
\begin{eqnarray}\label{25d}
\nonumber\gamma_{1}&=&\frac{\alpha_{1}}{3n-1}+\frac{n-3}{n\beta}
-\frac{2(3n-1)^{2}}{n^{2}\beta^{2}\alpha_{1}},\\
\nonumber\gamma_{2}&=&\frac{n(n-3)\beta\alpha_{1}}{(3n-1)^{2}},~~~\gamma_{3}=
\omega_{0}\beta^{2}n^{2}a_{0}^{\frac{2}{n}},~~~\gamma_{4}=m+2-\frac{2}
{n\beta},\\
\gamma_{5}&=&-2 \kappa^2 \rho_{0}a_{0}^{3(1+w)},~~~\gamma_{6}=-\frac{3}{\beta}.
\end{eqnarray}
Inserting (\ref{32}) in the energy conditions (\ref{9})-(\ref{12}) we can
find the energy conditions for this model. Here we are discussing the
validity of NEC and WEC for different values of $\beta$, $m$ and $t$ by
fixing $n$ and $\alpha_{i}$'s where $i=1,2$. Starting with $\alpha_{1}$ and
$\alpha_{2}$ both as positive, WEC is valid for all values of $m$ and
$\beta\neq0$ with $n=3$ while NEC is valid for $n=3$ with $\beta\leq-2$,
$m\geq0$ and $t\geq1.03$. Now taking $\alpha_{1}$ as negative and
$\alpha_{2}$ as positive, WEC is valid for $m\geq0$ with $n=3$,
$\beta\geq2.6$ $\&$ $t\geq0.65$ and for $m\leq-2$ it is valid for $n=3$ with
$\beta\geq22.5$. For this choice of $\alpha_i$'s NEC is valid for $n>1$ with
$\beta>1$, $m\leq-5$ $\&$ $t\geq1.05$. Next we are taking $\alpha_{1}$ as
positive and $\alpha_{2}$ as negative, here WEC is valid for $n=3$ with $(i)$
$\beta\geq2.7$ , $m\geq0$ $\&$ $t\geq0.65$ and with $(ii)$ $\beta\leq-2$,
$m\leq-5.5$ and $t\geq0.65$. NEC is valid for $n=3$ and for all values of $m$
and $\beta$ except $\beta=0$. If we take $\alpha_{1}$ and $\alpha_{2}$ both
as negative, both WEC and NEC are valid for all values of $m$ and
$\beta\neq0$ with $n=3$.

\section{Energy Conditions for Some known Models}

To present how these energy conditions apply limits on $f(R, Y, \phi)$
gravity, we have also considered some well-known functions in the following
discussion.

\subsection{$f(R,\phi)$ Models}

Here, we present $f(R,Y,\phi)$ gravity models which does not involve variation
with respect to $Y$ and corresponds to $f(R,\phi)$ gravity. We present the
energy constraints for the following models\\
$1.~f(R,\phi)=\frac{R-2\Lambda(1-e^{b\phi\kappa^{3}R})}{\kappa^{2}}$\\
$2.~f(R,\phi)=R\left(\frac{\omega_{0}\beta^{2}n^{2}a_{0}^{2/n}(mn\beta+2n\beta
    +6n-2)}{mn\beta+2n\beta-2}\right)\phi^{m+2-\frac{2}{n\beta}}$\\
$3.~f(R,\phi)=R(1+\xi\kappa^{2}\phi^{2})$\\
$4.~f(R,\phi)=\phi(R+\alpha R^{2})$\\
For these models we explore the energy constraints in the background of power law solutions with $n>1$
favoring the current accelerated cosmic expansion.

\subsubsection{Model-I}

In \cite{35}, Myrzakulov et al. discussed the inflation in $f(R,\phi)$ theories
by analyzing the spectral index and tensor-to-scalar ratio and found results in agreement
with the recent observational data. In our paper, we have selected the following $f(R,\phi)$ model \cite{35}
\begin{equation}\label{33}
f(R,\phi)=\frac{R-2\Lambda(1-e^{b\phi\kappa^{3}R})}{\kappa^{2}},~~
\omega(\phi)=1.
\end{equation}
where $\kappa^{3}$ is introduced for dimensional reasons and $b$ is a
dimensionless number of order unity.

Introducing this model in the energy conditions (\ref{9})-(\ref{12})
along with Eqs.(\ref{27}), (\ref{35}) and (\ref{25}), we find the
following constraints
\begin{eqnarray}
\nonumber&&\textbf{NEC:}~~\kappa^2(\rho_{0}t^{-3n(1+w)}+p)-\beta^{2}H^{2}
a_{0}^{2\beta}t^{2n\beta}+2\Lambda b \kappa \beta H^{2}a_{0}^{\beta}t^{n\beta}
\left(\beta-q-2\right)\\
\nonumber&&\times e^{-6b\phi\kappa^{3}(1-q)H^{2}}-12\Lambda b^{2}\kappa^{4}
H^{4}a_{0}^{2\beta}t^{2n\beta}\bigg[\left(s+q^{2}+8q+6\right)+\beta\left(1-
q\right)\times\\
\nonumber&&\left(\beta-1-q\right)+2\beta\left(j-q-2\right)+2\beta^{2}\left(1
-q\right)-\left(j-q-2\right)-\beta\left(1-q\right)\bigg]\times\\
\nonumber&& e^{-6b\phi\kappa^{3}(1-q)H^{2}}+72\Lambda b^{3}\kappa^{7}H^{6}
a_{0}^{3\beta}t^{3n\beta}\bigg[\beta^{2}\left(1-q\right)^{2}+2\beta\left(1-
q\right)\left(j-q-2\right)\bigg]\\
&&\times e^{-6b\phi\kappa^{3}(1-q)H^{2}}\geq0,\label{33a}\\
&&\textbf{WEC:}~~\nonumber\kappa^2\rho_{0}t^{-3n(1+w)}+\frac{1}{\kappa^{2}}\left
[3(1-q)H^{2}-\Lambda\left(1-e^{-6b\phi\kappa^{3}(1-q)H^{2}}\right)\right]
-\frac{1}{2}\beta^{2}\\
&&\nonumber\times H^{2}a_{0}^{2\beta}t^{2n\beta}-\Lambda b\kappa H^{2}
a_{0}^{\beta}t^{n\beta} \left(\beta-6+6q\right)e^{-6b\phi\kappa^{3}(1-q)H^{2}}
+36\Lambda b^{2}\kappa^{4}H^{4}\times\\
&& a_{0}^{2\beta}t^{2n\beta}\left(\beta(1-q)+(j-q-2)\right)e^{-6b\phi\kappa^{3}
(1-q)H^{2}}\geq0,\label{33b}\\
&&\textbf{SEC:}~~\nonumber\kappa^2(\rho_{0}t^{-3n(1+w)}+3p)+\frac{2\Lambda}
{\kappa^{2}}\left(1-e^{-6b\phi\kappa^{3}(1-q)H^{2}}\right)-2\beta^{2}H^{2}
a_{0}^{2\beta}t^{2n\beta}\\
&&\nonumber +6\Lambda b \kappa H^{2}a_{0}^{2\beta}t^{2n\beta}\bigg[\beta(\beta
-1-q)+\beta-2(1-q)\bigg]e^{-6b\phi\kappa^{3}(1-q)H^{2}}-36\Lambda b^{2}\times\\
&&\nonumber \kappa^{4}H^{4}a_{0}^{2\beta}t^{2n\beta}\bigg[(s+q^{2}+8q+6)+\beta
(1-q)(\beta-1-q)+4\beta(j-q-2)\\
&&\nonumber+2\beta^{2}(1-q)+(j-q-2)+\beta(1-q)\bigg]e^{-6b\phi\kappa^{3}(1-q)
H^{2}}+216\Lambda b^{3}\kappa^{7}H^{6}\times\\
&&\nonumber a_{0}^{3\beta}t^{3n\beta}\bigg[\beta^{2}(1-q)^{2}+2\beta(1-q)
(j-q-2)+(j-q-2)^{2} \bigg]\times\\
&&e^{-6b\phi\kappa^{3}(1-q)H^{2}}\geq0,\label{33c}
\end{eqnarray}
\begin{eqnarray}
&&\textbf{DEC:}~~\nonumber\kappa^2(\rho_{0}t^{-3n(1+w)}-p)-\frac{2\Lambda}
{\kappa^{2}}\left(1-e^{-6b\phi\kappa^{3}(1-q)H^{2}}\right)-2\Lambda b \kappa
H^{2}a_{0}^{\beta}t^{n\beta}\times\\
&&\nonumber\bigg[\beta(\beta-1-q)+5\beta-6(1-q)\bigg]e^{-6b\phi\kappa^{3}(1-q)
H^{2}}+12\Lambda b^{2}\kappa^{4}H^{4}a_{0}^{2\beta}t^{2n\beta}\times\\
&&\nonumber \bigg[(s+q^{2}+8q+6)+\beta(\beta-1-q)(1-q)+4\beta(j-q-2)
+2\beta^{2}(1-q)\\
&&\nonumber+5(j-q-2)+5\beta(1-q)\bigg] e^{-6b\phi\kappa^{3}(1-q)H^{2}}-72
\Lambda b^{3}\kappa^{7}H^{6}a_{0}^{3\beta}t^{3n\beta}\bigg[\beta^{2}(1-q)^{2}\\
&&\nonumber+2\beta(1-q)(j-q-2)+(j-q-2)^{2}\bigg]e^{-6b\phi\kappa^{3}(1-q)H^{2}}
\geq0.\label{33d}
\end{eqnarray}

Here, we are left with four parameters $b$, $\beta$, $n$ and $t$ and we constrain
these according to WEC and NEC. Starting with $b\geq0$, NEC is valid for $n>1$
with $\beta\leq-1.5$ whereas WEC is only valid for $b=0$ with $n>1$, $\beta\leq0$
and $t\geq1.1$. Moreover, for $b<0$ with $n>1$, NEC and WEC are valid for all
values of $\beta$. In Figure 2, we show the plot of NEC for this model verses the
parameters $m$, $\beta$ and $t$ by fixing $n>1$.
\begin{figure}
\centering \epsfig{file=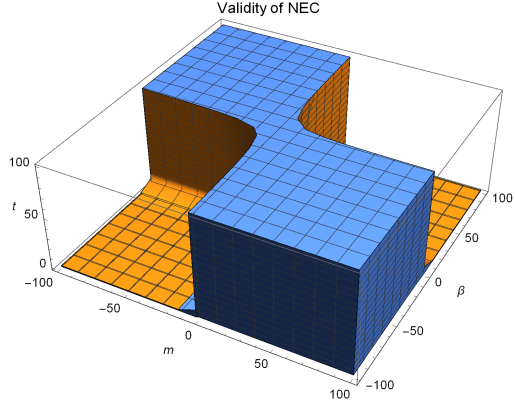, width=.49\linewidth,
height=2.1in} \caption{Plot of NEC for Model-II versus the parameters $m$,
$\beta$ and $t$ with $n=1.1$.}
\end{figure}

\subsubsection{Model-II}

Here, we have formulated a specific model in this theory using the form
$f(R,\phi)=Rf(\phi)$. We have calculated $f(\phi)$ from Klein-Gordon
equation by using $\omega(\phi)=\omega_{0}\phi^m$ and $\phi=a(t)^\beta$
given in $\cite{28}$,
\begin{equation}\label{34a}
2\omega(\phi)\Box \phi+ \omega_{\phi}(\phi)\phi_{;\alpha} \phi^{;\alpha}- f_{\phi}=0.
\end{equation}
In this regard, we find the following expression
\begin{equation}\label{34}
f(R,\phi)=R\left(\frac{\omega_{0}\beta^{2}n^{2}a_{0}^{2/n}(mn\beta+2n\beta
+6n-2)}{mn\beta+2n\beta-2} \right)\phi^{m+2-\frac{2}{n\beta}},
\end{equation}
where $\omega_{0}$ and $a_{0}$ are constants. Using this model in the
energy conditions (\ref{9})-(\ref{12}) along with Eqs.(\ref{27}), (\ref{35})
and (\ref{25}) we have energy conditions,
\begin{eqnarray}
&&\textbf{NEC:}~~\nonumber\kappa^2\rho_{0}t^{-3n(1+w)}+\kappa^2 p-\omega_{0}\beta^{2}
H^{2}a_{0}^{(m+2)\beta}t^{(m+2)n\beta}+\omega_{0}\beta^{2} H^{2}a_{0}^{(m+2)
\beta}\\
&&\nonumber\left\{(m+2)n\beta+2(3n-1)\right\}\left\{(m+2)n\beta-2(n+1)-nq
\right\}\\
&&\times t^{(mn\beta+2n\beta-2)}\geq0,\label{34a}\\
&&\textbf{WEC:}~~\nonumber\kappa^2\rho_{0}t^{-3n(1+w)}-\frac{1}{2}\omega_{0}
\beta^{2}H^{2}a_{0}^{(m+2)\beta}t^{(m+2)n\beta}-3\omega_{0}n\beta^{2} H^{2}
a_{0}^{(m+2)\beta}\\
&&\times \left\{(m+2)n\beta+2(3n-1)\right\}t^{mn\beta+2n\beta-2}\geq0,
\label{34b}\\
&&\textbf{SEC:}~~\nonumber\kappa^2(\rho_{0}t^{-3n(1+w)}+3p)-2\omega_{0}\beta^{2}
H^{2}a_{0}^{(m+2)\beta}t^{(m+2)n\beta}+3\omega_{0}\beta^{2} H^{2}
a_{0}^{(m+2)\beta}\\
&&\left\{(m+2)n\beta+2(3n-1)\right\}t^{(mn\beta+2n\beta-2)}\left\{(m\beta-q)n
+2(n\beta-1)\right\}\geq0,\label{34}\\
&&\textbf{DEC:}~~\nonumber\kappa^2(\rho_{0}t^{-3n(1+w)}-p)+\omega_{0}\beta^{2}
H^{2}a_{0}^{(m+2)\beta}\left\{(m+2)n\beta+2(3n-1)\right\}\\
&&\left\{n(q-m\beta)+2(1-2n-n\beta)\right\}t^{mn\beta+2n\beta-2)\beta}\geq0.
\label{34d}
\end{eqnarray}

We examine the NEC and WEC against the parameters $\beta$, $n$, $m$ and $t$.
We find that WEC can be satisfied for all values of $m$ and $\beta$ only if
$t\geq1.3$ while the validity of NEC requires; $(i)$ $m\geq0$ with
$\beta\leq0$ $\&$ $t\geq1.5$ $(ii)$ $m<-2$ with $\beta\geq0$ $\&$ $t\geq1.2$.

\subsubsection{Model-III}

In this case we present the energy constraints for the following model \cite{36}
\begin{equation}\label{40}
f(R,\phi)=R(1+\xi\kappa^{2}\phi^{2}),
\end{equation}
where $\xi$ is the coupling constant. Recently, this model has been
employed to discuss the cosmological perturbations for non-minimally
coupled scalar field dark energy in both metric and Palatini
formalisms. The interaction has been analyzed depending on the
coupling constant. Using this model in the energy conditions
(\ref{9})-(\ref{12}) along with Eqs.(\ref{27}), (\ref{35}) and
(\ref{25}) we get,\\
\begin{eqnarray}
&&\textbf{NEC:}~~\nonumber\kappa^2(\rho_{0}t^{-3n(1+w)}+p)-\omega_{0}\beta^{2}H^{2}
a_{0}^{(m+2)\beta}t^{(m+2)n\beta}+2\beta\xi\kappa^{2} H^{2}a_{0}^{2\beta}
t^{2n\beta}\\
&&\times(\beta-1-q)+2\beta^{2}\xi\kappa^{2}H^{2} a_{0}^{2\beta}t^{2n\beta}
-2\beta\xi\kappa^{2}H^{2}a_{0}^{2\beta}t^{2n\beta}\geq0,\label{40a}\\
&&\nonumber\textbf{WEC:}~~\kappa^2\rho_{0}t^{-3n(1+w)}-\frac{1}{2}\omega_{0}
\beta^{2}H^{2}a_{0}^{(m+2)\beta}t^{(m+2)n\beta}-6\beta H^{2}\xi\kappa^{2}\times\\
&&a_{0}^{2\beta}t^{2n\beta}\geq0,\label{40b}\\
&&\textbf{SEC:}~~\nonumber\kappa^2(\rho_{0}t^{-3n(1+w)}+3p)-2\omega_{0}\beta^{2}
H^{2}a_{0}^{(m+2)\beta}t^{(m+2)n\beta}+6\xi\kappa^{2}\beta H^{2}a_{0}^{2\beta}\\
&&\times t^{2n\beta}+6\xi\kappa^{2}\beta H^{2}(\beta-1-q)a_{0}^{2\beta}
t^{2n\beta}+6\xi\kappa^{2}\beta^{2} H^{2}a_{0}^{2\beta}t^{2n\beta}\geq0,
\label{40c}\\
&&\textbf{DEC:}~~\nonumber\kappa^2(\rho_{0}t^{-3n(1+w)}-p)-10\beta H^{2}\xi
\kappa^{2}a_{0}^{2\beta}t^{2n\beta}-2\beta H^{2}\xi\kappa^{2}(\beta-1-q)\\
&&\times a_{0}^{2\beta}t^{2n\beta}-2\beta^{2} H^{2}\xi\kappa^{2}a_{0}^{2\beta}
t^{2n\beta}\geq0.\label{40d}
\end{eqnarray}
We intend to discuss the NEC, WEC and constrain the parameters like
$\beta$, $\xi$, $n$, $m$ and $t$. Here, we develop three cases
depending on choice of scalar field power $m$. Starting with $m>0$
with $n>1$, NEC is valid for all values of $\xi$ with
$\beta\leq-3.7$ $\&$ $t\geq3$ and WEC is valid for all values of
$\xi$ with $\beta\leq-3.4$ $\&$ $t\geq2.8$. Now taking $m<0$ with
$n>1$, for $\beta\leq-3.7$ NEC is valid for all values of $\xi$ with
$t\geq3.1$ and for $\beta>0$ it is valid for all values of $t$ with
$\xi>0$. For $\beta\leq-3.4$ WEC is valid for all values of $\xi$
with $t\geq2.8$ and for $\beta\geq0$ WEC is valid for all values of
$t$ with $\xi\leq0$. Taking $m=0$ with $n>1$, WEC is valid in two
regions $(i)$ $\xi\leq-8.35$ with $\beta\geq0$ $(ii)$ for all $\xi$
with $\beta\leq-3.4$ and $t\geq2.8$. Similarly, NEC is satisfied for
$(i)$ $\beta\geq0$ with $\xi\geq0.28$ $(ii)$ for all $\xi$ with.
$\beta\leq-3.7$ and $t\geq3$

\subsubsection{Model-IV}

Bahamonde, S. et al has used the expression $f(R)$ \cite{37}
\begin{equation}
f(R,\phi)=\phi(R+\alpha R^{2}),
\end{equation}
where $\alpha$ is a constant with suitable dimensions. This gravitational
action is very familiar in the text as it is able to reproduce inflation.
Inserting this model in the energy conditions (\ref{9})-(\ref{12}) along with
Eqs.(\ref{27}), (\ref{35}) and (\ref{25}) we have energy conditions,\\
\begin{eqnarray}
&&\textbf{NEC:}~~\nonumber\kappa^2(\rho_{0}t^{-3n(1+w)}+p)-\omega_{0}\beta^{2}
H^{2}a_{0}^{(m+2)\beta}t^{(m+2)n\beta}-24\alpha \beta H^{4} a_{0}^{\beta}
t^{n\beta}\times\\
&&\nonumber(j-q-2)-12\alpha H^{4} (s+q^{2}+8q+6)a_{0}^{\beta}t^{(n\beta}
+\beta(\beta-1-q) H^{2}a_{0}^{\beta}t^{n\beta}\\
&&\nonumber -12\alpha \beta H^{4}(\beta-1-q)(1-q) a_{0}^{\beta}t^{n\beta}+12
\alpha H^{4}(j-q-2)a_{0}^{\beta}t^{n\beta}-\beta H^{2} a_{0}^{\beta}t^{n\beta}\\
&&+12\alpha \beta H^{4}(1-q)a_{0}^{\beta}t^{n\beta}\geq0,\\
&&\textbf{WEC:}~~\nonumber\kappa^2\rho_{0}t^{-3n(1+w)}-\frac{1}{2}\omega_{0}\beta
H a_{0}^{(m+2)\beta}t^{(m+2)n\beta}-18\alpha H^{4}(1-q)^{2}a_{0}^{\beta}
t^{n\beta}\\
&&+36\alpha H^{4}(j-q-2)a_{0}^{\beta}t^{n\beta}-3\beta H^{2}a_{0}^{\beta}
t^{n\beta}+36\alpha \beta H^{4}(1-q)a_{0}^{\beta}t^{n\beta}\geq0,\\
&&\textbf{SEC:}~~\nonumber\kappa^2(\rho_{0}t^{-3n(1+w)}+3p)-2\omega_{0}\beta^{2}
H^{2}a_{0}^{(m+2)\beta}t^{(m+2)n\beta}+36\alpha H^{4}(1-q)^{2}\\
&&\nonumber\times a_{0}^{\beta}t^{n\beta}-36\alpha H^{4}(j-q-2)a_{0}^{\beta}
t^{n\beta}+3\beta H^{2}a_{0}^{\beta}t^{n\beta}-36\alpha H^{4}(s+q^{2}+8q\\
&&\nonumber +6)a_{0}^{\beta}t^{n\beta}-36\alpha \beta H^{4}a_{0}^{\beta}
t^{n\beta}+3\beta H^{2}(\beta-1-q)a_{0}^{\beta}t^{n\beta}-72\alpha\beta H^{4}\\
&&\times (j-q-2)a_{0}^{\beta}t^{n\beta}-36\alpha \beta H^{4}(\beta-1-q)(1-q)
a_{0}^{\beta}t^{n\beta}\geq0,\\
&&\textbf{DEC:}~~\nonumber\kappa^2(\rho_{0}t^{-3n(1+w)}-p)+36\alpha H^{4}(1-q)^{2}
a_{0}^{\beta}t^{n\beta}+60\alpha H^{4}(j-q-2)\\
&&\nonumber\times a_{0}^{\beta}t^{n\beta}-5\beta H^{2}a_{0}^{\beta}t^{n\beta}
+60\alpha \beta H^{4}(1-q)a_{0}^{\beta}t^{n\beta}+24\alpha \beta H^{4}(j-q-2)\\
&&\nonumber \times a_{0}^{\beta}t^{n\beta}+12\alpha H^{4}(s+q^{2}+8q+6)
a_{0}^{\beta}t^{n\beta}-\beta H^{2}(\beta-1-q)a_{0}^{\beta}t^{n\beta}\\
&&+12\alpha \beta H^{4}(\beta-1-q)(1-q)a_{0}^{\beta}t^{n\beta}\geq0.
\end{eqnarray}

We are considering here NEC and WEC and check their validity for
different values of $\beta$, $\alpha$, $n$, $m$ and $t$. Following
the previous case we vary the coupling parameter $\alpha$ and set
the other parameters for the validity of WEC and NEC. If $\alpha>0$
with $n>1$, then WEC can be met in two regions namely,
($\beta\geq0$, $m\leq-1$ with $t\geq1$) and (for all values of $m$
with $\beta\leq-9$ and $t\geq6$). Now taking $\alpha<0$ with $n>1$,
WEC is valid for all $m$ with $\beta\leq0$, and NEC is valid if
$\beta\leq-0.7$ with $m\geq0$ and $t\geq1$ and for $\beta\geq0.85$
with $m\leq-1$ and $t\geq1$. Taking $\alpha=0$ with $n>1$, for
$\beta\geq0$ NEC is valid for $m\leq-1.05$ with $t>1.01$ and for
$\beta\leq0$ it is valid for $m\geq0$ with $t\geq1$. WEC is valid
for all values of $m$ with $\beta\leq0$.

\section{Conclusion}

Scalar tensor theories of gravity are very useful to discuss accelerated
cosmic expansion and to predict the universe destiny. One of more
general modified gravity is, $f(R,R_{\mu\nu}R^{\mu\nu},\phi)$ which
include the contraction of Ricci tensors $Y=R_{\mu\nu}R^{\mu\nu}$ and scalar field $\phi$.
In this paper, we have applied the reconstruction programme to $f(R,R_{\mu\nu}R^{\mu\nu}, \phi)$.
The action (\ref{18}) in original and specific forms $f(R,\phi)$, $f(Y, \phi)$ is reconstructed for
some well-known solutions in FRW background. The existence of dS solutions has been investigated in modified
theories \cite{23*}. Here, we have developed multiple dS solutions which can handy in explaining the differenr
cosmic phenomena. In de-Sitter universe,
we have constructed the more general case $f(R,Y,\phi)$ and establish
$f(R,\phi)$ considering the function independent of $Y$ and $f(Y,\phi)$
by taking function independent of $R$. The power law expansion history has also been reconstructed in this
modified theory for both general as well as particular form of the
action (\ref{18}). These solutions explain the matter/radiation
dominated phase that connects with the accelerating epoch. The
$f(R,R_{\mu\nu}R^{\mu\nu},\phi)$ model can also be reconstructed which will reproduce the
crossing of phantom divide exhibiting the superaccelerated expansion
of the universe.

Lagrangian of $f(R,R_{\mu\nu}R^{\mu\nu},\phi)$ gravity is more comprehensive
implying that different functional forms of $f$ can be suggested. The
versatility in Lagrangian raises the question how to constrain such theory on
physical grounds. In this paper, we have developed some constraints on general
as well as specific forms of $f(R,T,R_{\mu\nu}T^{\mu\nu})$ gravity by examining
the respective energy conditions. The energy conditions are also developed in terms of
deceleration $q$, jerk $j$, and snap $s$ parameters. To illustrate how these conditions can constrain the
$f(R,R_{\mu\nu}R^{\mu\nu},\phi)$ gravity, we have explored the free parameters in reconstructed
and well known models. In general dS case $f(R,Y,\phi)$ energy conditions are depending
on six parameters $\beta$, $m$, $t$ and $\alpha_{i}$'s where $i=1,2,3$.
In this procedure we have fixed the $\alpha_{i}$'s and observe the
feasible region by varying the other parameters.

In dS $f(R,\phi)$ and $f(Y,\phi)$ models, the NEC depend on five parameters $\alpha_{1}$,
$\alpha_{2}$, $\beta$, $m$ $\&$ $t$ and WEC depend only on three
parameters $\alpha_{1}$, $\alpha_{2}$ $\&$ $t$. In case of NEC we have
fixed $\alpha_{1}$ and $\alpha_{2}$ and find the constraints on the
other parameters. In WEC we are changing $\alpha_{1}$ and explore the
possible ranges on $\alpha_{2}$ and $t$. For power law $f(R,\phi)$
and $f(Y,\phi)$ models, functions depend on six parameters $\alpha_{1}$,
$\alpha_{2}$, $\beta$, $m$, $n$ and $t$. In power law case we have $n>1$,
and varying $\alpha_{1}$, $\alpha_{2}$ we have analyzed the viable
constraints on $\beta$, $m$ and $t$. Further more we have considered three
particular forms of $f(R,Y,\phi)$ gravity taking function independent of $Y$,
i.e., $f(R,\phi)$, $Rf(\phi)$, $\phi f(R)$ from which we can deeply understand
the applications of energy conditions. Model-I is a function of four parameters
$b$, $\beta$, $n$ and $t$, we have checked the validity of NEC and WEC by
varying $b$. Model-II is depending on $\beta$, $m$, $n$ and $t$, for $n>1$
we have explored the viability of other parameters. Next in model-III we have
five parameters $\beta$, $\xi$, $n$, $m$ and $t$, for $n>1$ we have find the
feasible constraints on other parameters by fixing $m$. In model-IV the
conditions are depending on five parameters $\beta$, $\alpha$, $n$, $m$ and $t$.
we have $n>1$ and varying $\beta$ we examined the possible regions for the
other parameters.

Finally, we generally discuss the variations of parameters involved in power
law solutions and scalar field coupling function, denoted by $m$ and $n$
respectively. In de-Sitter models we have examined that the more general case
$f(R,Y,\phi)$ is more effective as compared to $f(R,\phi)$ and $f(Y,\phi)$
models since in general case one can specify the parameters in more
comprehensive way. In all cases of de-Sitter models, WEC is valid for all $m$
and NEC is valid if ($m\geq1$ $\&$ $m\leq-5$). In power law case $f(R,\phi)$,
for both NEC and WEC $n$ has a fixed value $n=3$ and $m$ has variations
($m\geq0$ $\&$ $m\leq-5.5$). For $f(Y,\phi)$ case we have ($n\geq2.3$ with
$m\geq4$, $m\leq-1$) for WEC and for NEC we have $n\geq2$ with ($m\geq0$,
$m\leq-4$). In other known $f(R,\phi)$ models, the validity of these
conditions require $n>1$ with ($m\geq0$, $m\leq-2$).

\end{document}